\documentclass[nofootinbib]{revtex4}

\usepackage{dsfont}
\usepackage{graphicx} 
\usepackage{epstopdf} 
\usepackage{slashed}
\usepackage{color}
\usepackage{subfigure}

\newcommand{\definmath}[2] {\def#1{\ifmmode#2\else$#2$\fi}}

\newcommand{\qed}{\nobreak \ifvmode \relax \else
      \ifdim\lastskip<1.5em \hskip-\lastskip
      \hskip1.5em plus0em minus0.5em \fi \nobreak
      \vrule height0.75em width0.5em depth0.25em\fi}

\newcommand{\hideIt}[1]{{}}

\newcommand{\mttwo}{M_{T2}}
\newcommand{\mct}{M_{CT}}
\newcommand{\hide}[1]{}

\newcommand{\eap}{{  ( \hat{\bf a} . {   \hat{\slashed{\bf p}}   }' )  }}
\newcommand{\ebp}{{  ( \hat{\bf b} . {   \hat{\slashed{\bf p}}   }' )  }}
\newcommand{\esigmap}{{  ( {\bf \sigma} . {   \hat{\slashed{\bf p}} }' )  }}
\newcommand{\edeltap}{{  ( {\bf \delta} . {   \hat{\slashed{\bf p}} }' )  }}
\newcommand{\sigmadotp}{{({\bf \sigma}.{ \hat{\slashed{\bf p}}}) }}
\newcommand{\deltadotp}{{({\bf \delta}.{ \hat{\slashed{\bf p}}}) }}
\newcommand{\ahdotph}{{(\hat{\bf a}.{ \hat{\slashed{\bf p}}}) }}
\newcommand{\bhdotph}{{(\hat{\bf b}.{ \hat{\slashed{\bf p}}}) }}

\newcommand{\eahbh}{{   ({\bf \hat  a}.{{\bf \hat b}'})    }}
\newcommand{\ahdotbh}{{   ({\bf \hat a}.{{\bf \hat  b}})    }}
\newcommand{\cosTheta}{{\cos\theta}}
\newcommand{\sinTheta}{{\sin\theta}}

\newcommand{\vecPtmiss}{{\slashed{\bf p}}}

\newcommand{\Kone}{{K_1}}
\newcommand{\Ks}{{K_s}}
\newcommand{\Kc}{{K_c}}
\newcommand{\Kss}{{K_{ss}}}
\newcommand{\Kcc}{{K_{cc}}}
\newcommand{\Kcs}{{K_{cs}}}
\newcommand{\AAA}{{A}}
\newcommand{\BB}{{B}}
\newcommand{\CC}{{C}}
\newcommand{\DD}{{D}}
\newcommand{\EE}{{E}}

\newcommand{\thenewvar}{{$\mttwo^{\rm approx}$}}

\newcommand{\eqref}[1]{{(\ref{#1})}}

\begin{document}   
\title{Properties of $\mttwo$ in the massless limit}
\date{\today}

\author{Colin H. Lally}
\email{lallyc@cranbrook.kent.sch.uk}
\affiliation{Cranbrook School, Waterloo Road, Kent, TN17 3JD, United Kingdom}
\author{Christopher G. Lester}
\email{lester@hep.phy.cam.ac.uk}
\affiliation{Cavendish Laboratory, Dept of Physics, JJ Thomson Avenue, Cambridge, CB3 0HE, United Kingdom}

\begin{abstract} 
Although numerical methods are required to evaluate the stransverse
mass, $\mttwo$, for {\em general} input momenta, non-numerical methods
have been proposed for some special classes of input momenta.  One
special case, considered in this note, is the so-called ``massless
limit'' in which all four daughter objects (comprising one invisible
particle and one visible system from each ``side'' of the event) have
zero mass.  This note establishes that it is possible to construct a
{\em stable and accurate} implementation for evaluating $\mttwo$ based
on an analytic expression valid in that massless limit.  Although this
implementation is found to have no significant speed improvements over
existing evaluation strategies, it leads to an unexpected by-product:
namely a secondary variable, ``\thenewvar'', that is found to be very
similar to $\mttwo$ for much of its input-space and yet is much faster
to calculate.  This is potentially of interest for hardware applications that
require fast estimation of a mass scale (or QCD background
discriminant) based on a hypothesis of pair production -- as might be
required by a high luminosity trigger for a search for pair production
of new massive states undergoing few subsequent decays (eg di-squark
or di-slepton production).  This is an application to which the
contransverse mass $M_{CT}$ has previously been well suited due to its
simplicity and ease of evaluation.  Though \thenewvar\ requires a quadratic root to be found, \thenewvar\ (like $M_{CT}$) does not require iteration to compute, and is found to perform
better then $M_{CT}$ in circumstances
in which the information from the missing transverse momentum (which
the former retains and the latter discards) is both reliable and 
useful.
%

\end{abstract}   
\maketitle 


\section{Introduction}

\label{sec:lalintro}

The stransverse mass, $\mttwo$, \cite{Lester:1999tx} is a kinematic
variable which is defined to be the maximal lower-bound on the
mass of a parent particle decaying to an invisible daughter of a known
(or hypothesised) mass and one or more visible daughter particles in
a hadron collider event, provided that there are two of these parents in the event.  Properties and generalisations of $\mttwo$ have been investigated extensively \cite{Allanach:2000kt,Barr:2002ex,Barr:2003rg,Lester:2007fq,Cho:2007qv,Gripaios:2007is,Barr:2007hy,Cho:2007dh,Ross:2007rm,Nojiri:2007pq,Tovey:2008ui,Cho:2008cu,Serna:2008zk,Barr:2008ba,Cho:2008tj,Burns:2008va,Barr:2008hv,Barr:2009mx,Barr:2009jv,Polesello:2009rn,Kim:2009si,Konar:2009wn,Konar:2009qr,Baringer:2011nh}.
Experimentally, $\mttwo$ has been used in searches for R-parity conserving supersymmetry signals \cite{daCosta:2011qk,Collaboration:2012ida,c:2012gg,cm:2012jx,Weber:2012fa}, Higgs searches (for example \cite{d0000:2012qq,Barr:2011he}) and in measurements of the top quark mass  \cite{Aaltonen:2009rm}.


Although its properties as a mass-scale identifier and QCD-background discriminator are well established, it is not possible {\em in general} to write down a closed-form analytic expression for $\mttwo$ or many of the variables to which it is related.  For some discussion on this point see \cite{Lester:2007fq,Burns:2008va,Cheng:2008hk}. To date therefore, $\mttwo$ and similar variables are generically evaluated using iterative numerical techniques, and code is available from various sources to do this \cite{oxbridgeStransverseMassLibrary,ucdWimpmassLibrary}. It has, however, been shown that in a wide class of special cases \cite{Lester:2011nj} analytic formulae for $\mttwo$ can be found. In one of these, specifically the so-called ``fully massless case'' in which the daughter particle masses are all assumed to be zero, $\mttwo$ may be written as a simple function of one of the two or four real roots of a quartic equation:
\begin{equation}
A s^4 + \BB s^3 + \CC s^2 + \DD s + \EE = 0\label{eq:quarticins} 
\end{equation}
whose coefficients are simple functions of the observed momenta.\footnote{Explicit expressions for the five coefficients appear later in the paper.}   On its own, this result is perhaps little more than a curiosity.  Quartic roots are not terribly pleasant things to work with -- so much so that they are frequently treated numerically even though analytic formulae for them exist.  As a result, very few implementations have made use of analytic solutions of equation (\ref{eq:quarticins}) given fears of instability, and it is even less clear that any useful insights are likely to emerge from pursuing this line further.  The idea that motivated this investigation was a desire to challenge that assumption --- a desire to see if anything useful was trapped inside equation (\ref{eq:quarticins}) after all.

The starting point, therefore, requires the construction and
validation of a stable method of calculating the roots of
equation (\ref{eq:quarticins}) using non-iterative analytic methods (described
in detail in the appendix).  The important part of that task is not the root finding
itself,\footnote{Nonetheless, ensuring {\em stability} in any root-finding code is non-trivial, and the appendix also details why the method employed here is regarded as being sufficiently stable.} but rather the understanding of the
characteristics of the input space that determine the characteristics
of the roots -- and in particular determine {\em which of those roots}
is the ``physical'' one that is ultimately used to calculate $\mttwo$.


This paper documents the insights that were gained regarding the geometric properties of $\mttwo$ found in the special case considered, and indicates how they could used to define an approximation to $\mttwo$, computable by a fixed-cost non-iterative method, possibly as an alternative to $\mct$ appropriate when the missing transverse momentum is reliable input, and where fast and reliable calculation times are required, for example in trigger routines.

\section{Revisiting the massless case}
\label{sec:findingroots}

Following the notation of \cite{Lester:2011nj} we revisit the
derivation of the expression for $\mttwo$ in the massless case that uses
(\ref{eq:quarticins}), with a view to highlighting the parts that are related to how the correct choices are made among those roots.



\begin{figure}
\includegraphics[width=0.3\columnwidth]{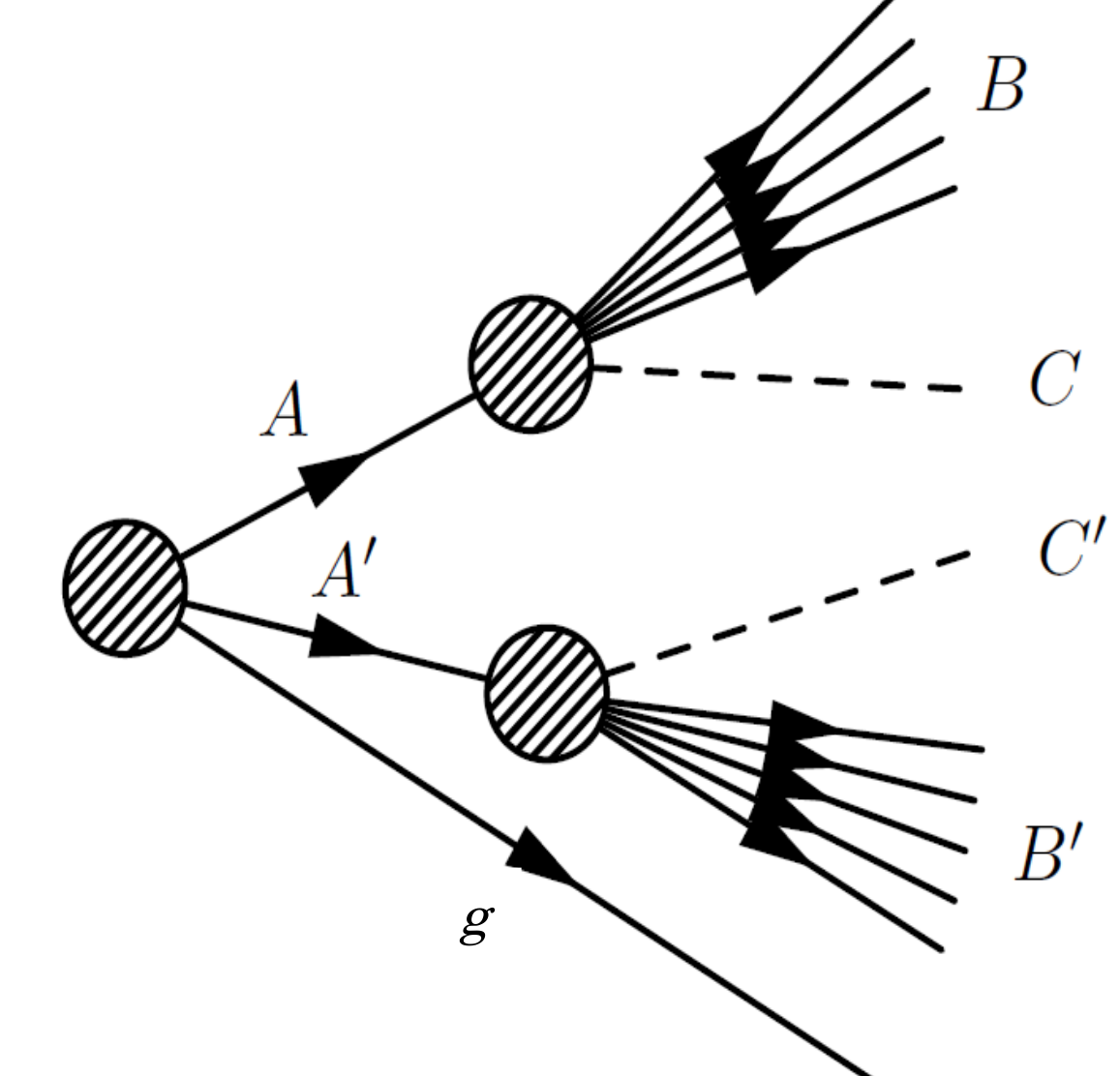}
\caption{The diagram shows an initial state (the left most blob) which produces a pair of parents ($A$ and $A'$) and some other source (or sources) of Upstream Transverse Momentum ($g$). Each visible particle decays to one or more visible particles ($B$) and an invisible particle ($C$).   \label{fig:typicalevent}}
\end{figure}

We begin with the hypothesis,  illustrated in figure~\ref{fig:typicalevent}, that two pair-produced parent particles ($A$, $A'$) each decay into a (group of) visible ($B$, $B'$) and invisible ($C$, $C'$) daughter particle(s). Each decay branch is referred to as one ``side'' of the event. The momenta of the visible decay products of sides 1 and 2 (assumed to have been reconstructed into two jets) are referred to as $a^\mu$ and $b^\mu$ respectively, and the hypothesised momenta of the invisible decay products as $p^\mu$ and $q^\mu$ respectively.  

We restrict ourselves to the transverse plane and so we view $a^\mu$ as consisting of three components: $a^\mu = (e_T, p_x, p_y)$, or $a^\mu =(e_T,{\bf a})$. Furthermore, as we assume a fully massless situation, then $a^2 (= e_T^2 - {\bf a}^2) = b^2 = p^2 = q^2 = 0$.


Of course the invisible momenta $p^\mu$ and $q^\mu$ will not be known; instead for each event the missing transverse momentum $\vecPtmiss$ will be obtained by adding together the visible transverse momenta, ${\bf a}$ and ${\bf b}$, and the ``upstream transverse momenta'' (UTM), denoted by ${\bf g}$. Thus we have the following relationships between the transverse momenta:
\begin{eqnarray}
	{\bf a}+{\bf b}+\vecPtmiss+{\bf g}={\bf 0}  \label{eq:longsplitting}\\
	{\bf p}+{\bf q}=\vecPtmiss \label{eq:shortsplitting}
\end{eqnarray}

Equation~\ref{eq:shortsplitting} is referred to as the ``momentum splitting condition''. Without loss of generality, we assume that $\vecPtmiss$ lies outside of the sector of the transverse plane spanned by positive multiples of $\bf a$ and $\bf  b$.\footnote{We remind the reader that in cases in which the $\vecPtmiss$ vector happens to lie ``between'' $\bf a$ and $\bf b$ then $\mttwo$ in the ``fully massless'' case must be identically zero (the so-called ``trivial zero'' of $\mttwo$) as we can satisfy the momentum splitting condition by taking $\bf p$ and $\bf q$ collinear with $\bf a$ and $\bf b$ respectively. This trivial case, in which $\mttwo$ is identically zero, is implicitely excluded from any of the subsequent discussion concerning methods of evaulating $\mttwo$, though it is always present in any actual evaluation or implementation. (See right-hand plot of figure~\ref{fig:backtobackutm} later for an example of this ``trivial zero'' scenario.)}   We also introduce the (2-D) unit vectors of the transverse momenta (actually these are the transverse \textit {velocities} of the particles as in the massless case $e_T = |{\bf p}|$):  $\hat {\bf a}$, $\hat {\bf b}$, $\hat {\bf p}$, $\hat {\bf q}$, $\hat \vecPtmiss$.

$\mttwo$ itself is defined as the minimum (over all hypothesised momenta $\bf p$ and $\bf q$ consistent with the momentum splitting condition) of the larger of the two transverse masses obtained from each ``side'' of the event. Symbolically:
\begin{equation}
\mttwo = \min_{{\bf p}+{\bf q} = \vecPtmiss} \left\{ \max \left[ M_T(a^\mu, {\bf p}), M_T(b^\mu, {\bf q})\right]\label{eq:mt2defles} \right\}
\end{equation}
For more details see \cite{Lester:1999tx}.  For now all we need to
note is that this is a min-max process, and in calculating $\mttwo$ our interest will be drawn primarily to the specific vectors ${\bf p}$ and ${\bf q}$ which minimise the max in the RHS of eq (\ref{eq:mt2defles}).  We will refer to these vectors as those which ``realise the extremum of the $\mttwo$ min-max process'', or sometimes simply as those vectors which ``lead to the $\mttwo$ solution'' {\it etc}.

A key property of the hypothesised vectors $\bf q$ and $\bf p$ that realise the extremum of the $\mttwo$ min-max process, is that the angle {\em between them} is the same as the angle between $\bf a$ and $\bf b$ \cite{Lester:2011nj}. As a consequence, $\hat {\bf p}$  and $\hat {\bf q}$ can be parameterised in terms of a single unknown parameter $\theta$ in the following way:
\begin{eqnarray}
\hat {\bf p} &=& \cosTheta {\hat{\bf b}} + \sinTheta {\hat{\bf b}'} \nonumber \\ 
\hat {\bf q} &=& \cosTheta {\hat{\bf a}} + \sinTheta {\hat{\bf a}'}\label{eq:parameterq}\label{eq:parameterp} 
\end{eqnarray}where the two vectors $\hat {\bf  a}'$ and $\hat {\bf b}'$ are defined to be the two vectors $\hat {\bf a}$ and $\hat {\bf b}$ rotated by +90 degrees in the transverse plane.  In effect, $\theta$ represents the angle by which $\hat {\bf a}$ and $\hat {\bf b}$ must be rotated by in order to arrive at the directions of $\hat {\bf q}$ and $\hat {\bf p}$ which lead to the $\mttwo$ solution.

As we have assumed that $\vecPtmiss$ lies outside of the sector of the transverse plane spanned by positive multiples of $\bf a$ and $\bf  b$, we are only interested in the so-called ``balanced'' momentum-hypothesis configurations where both sides of the event have the same transverse mass. In other words, $\bf p$ and $\bf q$ must also satisfy the following:\begin{equation}
2\left(\left|{\bf a}\right|\left|{\bf p}\right|-{\bf a}.{\bf p}\right)
=
2\left(\left|{\bf b}\right|\left|{\bf q}\right|-{\bf b}.{\bf q}\right),\label{eq:balancedeq}
\end{equation}
and when this contraint is met, each side of equation (\ref{eq:balancedeq}) will also be equal to the desired final value of $\mttwo^2$. Putting the ``opening-angle'' condition \eqref{eq:parameterp} together with the ``balanced hypothesis'' condition \eqref{eq:balancedeq} leads to the first of two useful results:
\begin{equation}
\left|{\bf a}\right|\left|{\bf p}\right|\left(1-\cosTheta \ahdotbh - \sinTheta \eahbh\right)
=\label{eq:parameterisedbalanced}
\left|{\bf b}\right|\left|{\bf q}\right|\left(1-\cosTheta \ahdotbh + \sinTheta \eahbh\right) 
\end{equation}
while taking the opening-angle condition \eqref{eq:parameterp} together with the momentum splitting condition \eqref{eq:shortsplitting} and a dot-product with $\vecPtmiss '$ ($\vecPtmiss$ rotated by +90 degrees in the transverse plane) leads to the second:
\begin{equation}
+\left|{\bf p}\right|\left(\cosTheta \ebp + \sinTheta \bhdotph\right)
= \label{eq:parameterisedshortsplitting}
-\left|{\bf q}\right|\left(\cosTheta \eap + \sinTheta \ahdotph\right).
\end{equation}
 
The quotient of these two constraints \eqref{eq:parameterisedbalanced} and \eqref{eq:parameterisedshortsplitting} eliminates $|{\bf p}|$ and $|{\bf q}|$ allowing $\theta$ to be found.  Indeed, it is precisely that quotient re-written as a quartic in terms of $s=\sin\theta$, that was referred to in equation \eqref{eq:quarticins} wherein, for completeness, we note that
\begin{eqnarray}
\AAA &=& \Kcs^2 + (\Kss - \Kcc)^2 \\
\BB &=&  2 (\Kcs \Kc + \Ks (\Kss - \Kcc))\\
\CC &=& \Ks^2 -\Kcs^2 + \Kc^2 + 2 (\Kss - \Kcc) (\Kone + \Kcc)\\
\DD &=&  2 (-\Kcs \Kc + \Ks (\Kone + \Kcc))\\
\EE &=& (\Kone - \Kc + \Kcc) (\Kone + \Kc + \Kcc)
\end{eqnarray}
using the definitions
\begin{eqnarray}
 \Kss &=& - \deltadotp  \eahbh \label{eq:firstkdef}\\
     \Kcc &=&  - \esigmap  \ahdotbh \\
     \Ks &=& \sigmadotp \\
     \Kc &=& \esigmap \\
     \Kcs &=& -\sigmadotp  \ahdotbh - \edeltap  \eahbh \\
     \Kone &=& 0  \label{eq:lastkdef} \\
\sigma &=& {\bf a} + {\bf b} \qquad\text{and} \\
\delta &=& {\bf a} - {\bf b}.
\end{eqnarray}

As before, momenta unit-vectors carry a ``hat'' whereas a ``prime'' (as in $\hat {\bf b}'$) signifies a rotation of the unit-vector through +90 degrees in the transverse plane.

\mbox{}

\begin{figure}
\mbox{\subfigure[ Typical event with two real roots]{\includegraphics[width=0.49\columnwidth]{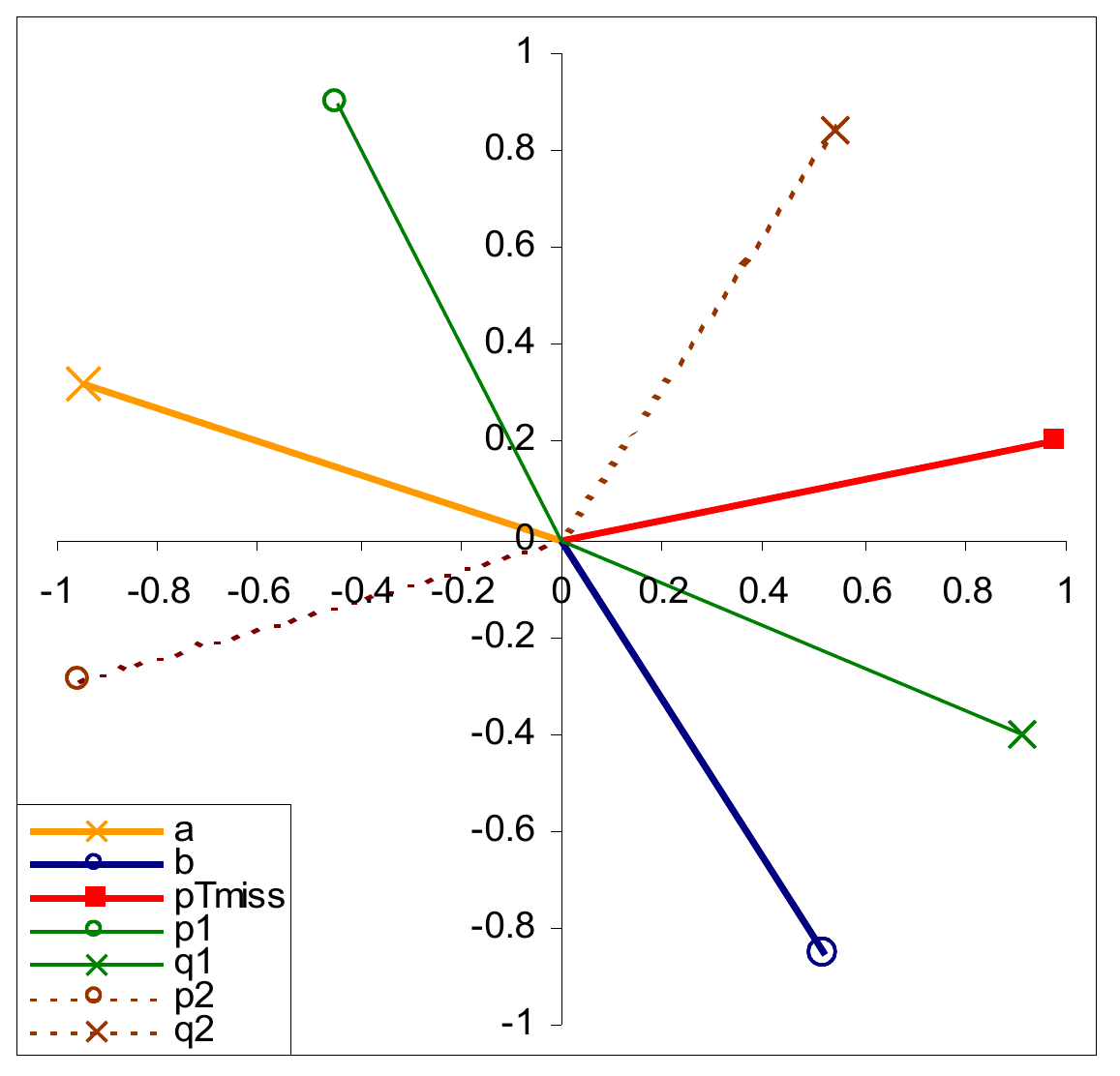}
\label{fig:2realroots}}
\subfigure[ Typical event with four real roots]{\includegraphics[width=0.49\columnwidth]{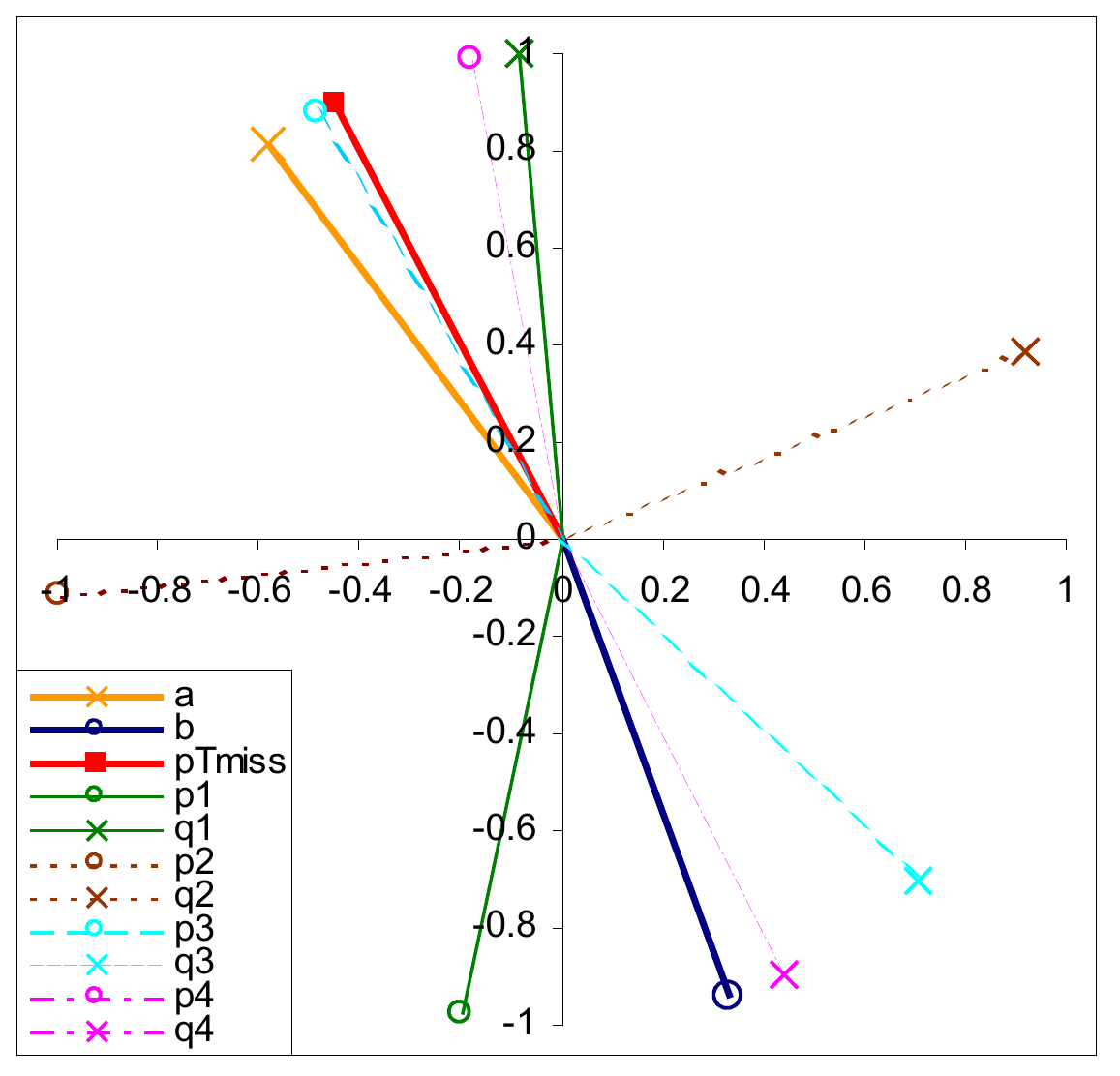}
\label{fig:4realroots}}}
\caption{
Unit-vector representation of momenta ${\bf a}$, ${\bf b}$ and $\vecPtmiss$, and candidate ${\bf p}$ and ${\bf q}$ values obtained from the roots of the quartic
\label{fig:examplerealroots}}
\end{figure}

The key to understanding which of the real roots (i.e.~always either two or four real values for $\sinTheta$) leads to the correct physical value for $\mttwo$, is this geometric picture of what we are doing to calculate $\mttwo$: rotating the hypothetical invisible ${\bf p}$ and ${\bf q}$ momenta away from the visible momenta ${\bf a}$ and ${\bf b}$ until both the ``balanced hypothesis'' constraint \eqref{eq:parameterisedbalanced} and the momentum splitting constraint \eqref{eq:parameterisedshortsplitting} are satisfied. What in effect this means is that the correct root must correspond to an angle that rotates ${\bf p}$ and ${\bf q}$ until one and only one of these vectors has been rotated ``past'' ${\vecPtmiss}$ thus ensuring constraint \eqref{eq:parameterisedshortsplitting} can be satisfied.\footnote{Note the rotation is defined to be always in the ``positive'', that is anticlockwise, sense.} Most of the time this rotation condition is sufficient to identify the correct physical root as most of the time we only have two real roots (the other two complex roots having been discarded as obviously unphysical).  Figure~\ref{fig:2realroots}, taken from a toy Monte Carlo event generator used to test the quartic root finder code (described later), shows a typical example of this situation where it is clear which of the two possible rotated pairs of vectors (i.e. the pair numbered ``1'' or ``2'') would satisfy the momentum splitting condition \eqref{eq:shortsplitting}\footnote{It is worth noting that the other real root (e.g. in this case the vector pair numbered ``2'' in figure~\ref{fig:2realroots}) corresponds to values for $|{\bf p}|$ and $|{\bf q}|$ which have \textit{negative} values. These still satisfy both the balanced hypothesis constraint \eqref{eq:parameterisedbalanced} and the momentum splitting constraint \eqref{eq:parameterisedshortsplitting}, but are clearly unphysical.}.

There are occasions where we have four real roots and more than one satisfies the rotation condition above. These all potentially correspond to physical values of $\mttwo$. But as $\mttwo$ by definition requires a {\em lower-bound} of the original parent masses, the final condition we require is that the correct root is the one whose corresponding angle $\theta$ ensures ${\bf p}$ and ${\bf q}$ are as parallel to ${\bf a}$ and ${\bf b}$ (respectively) as possible, as this will identify the lowest value for $\mttwo$ for the event in question (i.e. this by definition minimises both sides of equations~\eqref{eq:balancedeq} and~\eqref{eq:parameterisedbalanced}). Figure~\ref{fig:4realroots} shows a typical example of this situation, and here, although vector pairs ``1'', ``2'' and ``3'' correspond to possible physical rotations of ${\bf p}$ and ${\bf q}$\footnote{But not vector pair 4 as $\vecPtmiss$ does not lie ``in-between'' them and so they represent the unphysical solution for $\theta$.}, it is vector pair ``3'', i.e.~vectors ${\bf p}_3$ and ${\bf q}_3$, which are most ``parallel'' to ${\bf a}$ and ${\bf b}$ and so it is this pair which corresponds to the lowest $\mttwo$ value. 
Note that in very many cases, this is the same as saying it is the root whose rotation angle is as close to \textit{180 degrees} as possible - more on this point later.

\begin{figure}
\mbox{\subfigure[ Momenta distributions in the ``Low UTM'' scenario]{\includegraphics[width=0.46\columnwidth]{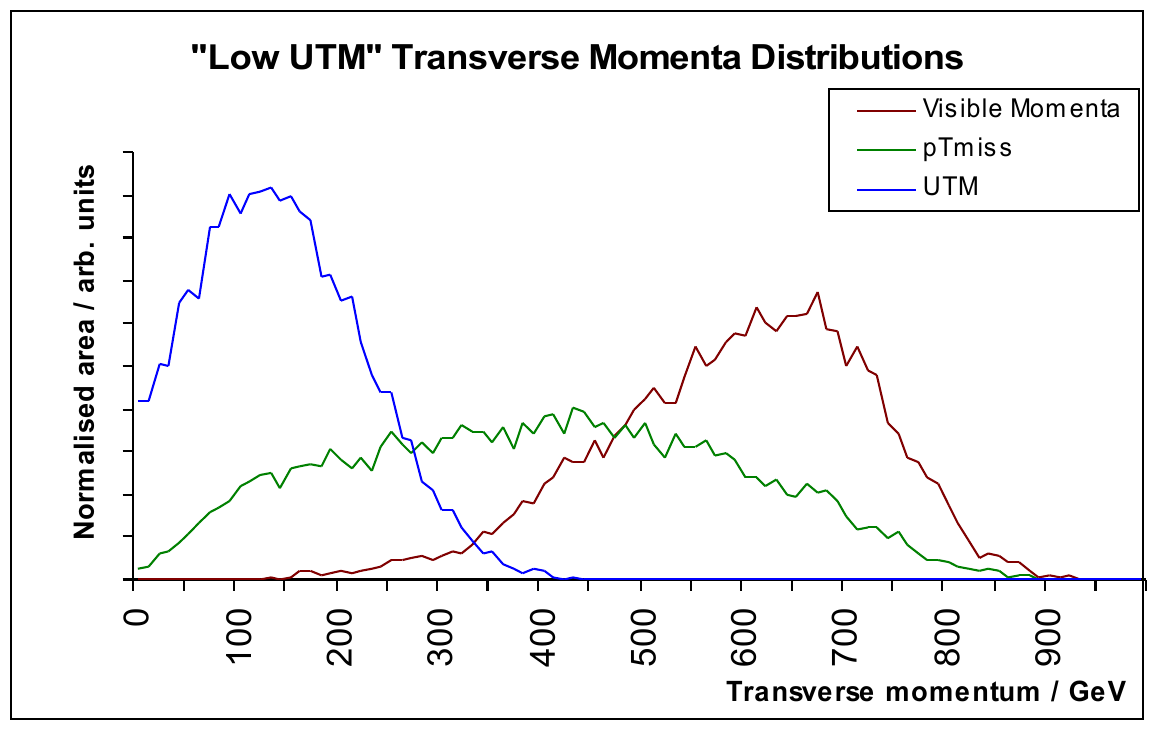}
\label{fig:lowUTMmomentadistributions}}
\subfigure[ Distribution of $\cosTheta$ in the ``Low UTM'' scenario]{\includegraphics[width=0.46\columnwidth]{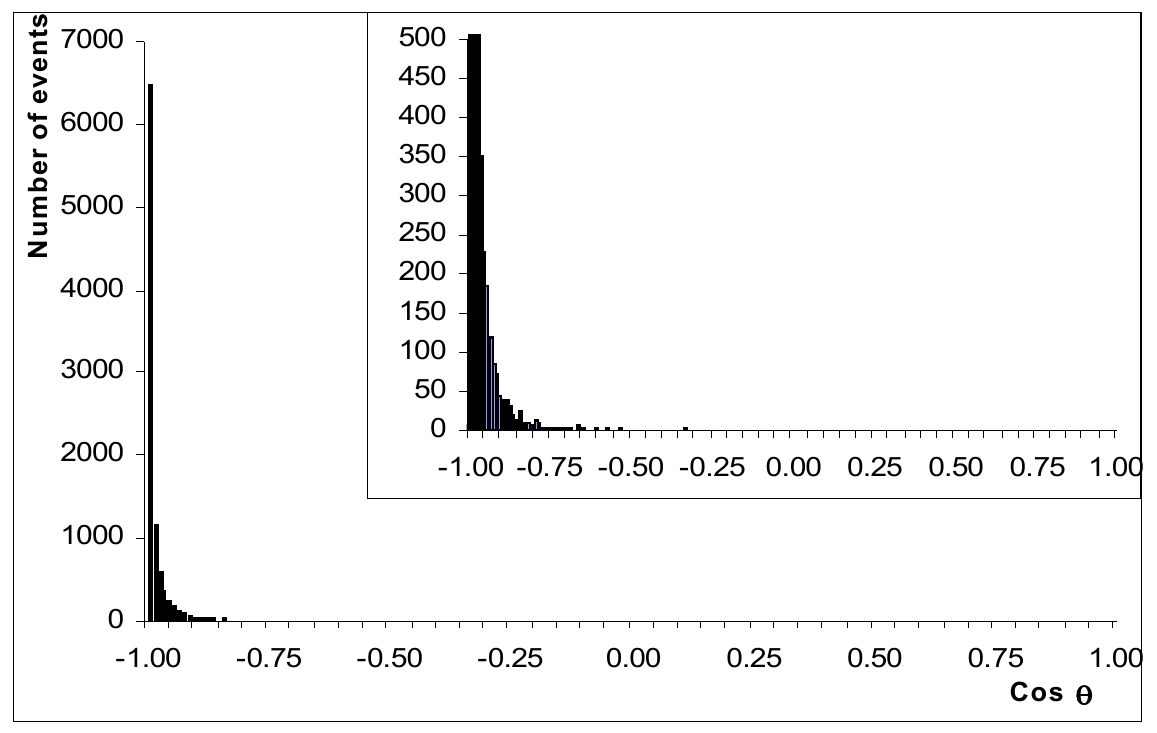}
\label{fig:lowUTMcosthetadistrib}}}
\caption{
Low UTM scenario input momenta and output rotation angles
\label{fig:lowUTMscenariofigs}}
\end{figure}

\section{An implementation}
\label{sec:performance}

Using a method due originally to Leonard Euler -- but expanded upon recently by Nickalls \cite{Nickalls:2009aa} whose work was especially useful in ensuring a stable root-finding method (detailed in the appendix) -- and the right root identification techniques outlined in the previous section, a {\tt C++} program was written to calculate $\mttwo$ in the fully massless case (the code can be found at \cite{oxbridgeStransverseMassLibrary}). 


\label{sec:analysis}

\subsection*{Testing and Accuracy}
In order to test this implementation, a simple ``toy'' Monte Carlo was used to generate signal-type events for the program to analyse. The toy Monte Carlo reproduces a simple event as depicted in figure~\ref{fig:typicalevent}; it begins with a normally distributed centre-of-mass ``rest mass'' which converts isotropically into two pair-produced parent particles (each possessing half the rest mass of the original ``rest mass''), which themselves decay into daughter visible and invisible particles (the latter's energy and momenta being represented by the missing energy and momenta of the event).  For the purposes of this analysis, the mean of the normally-distributed pair-produced parent masses was fixed at 750~GeV, and the visible and invisible decay particles were assumed to be massless. This leads to events with a wide range of momenta for the visible and invisible decay particles.  UTM in an event is introduced by including a normally distributed initial centre-of-mass energy (which can boost the momenta of the visible and invisible decay particles); adjusting this initial centre-of-mass energy distribution up or down produces events with more or less UTM.


The initial ``low-UTM'' run uses a mean centre-of-mass energy of 200~GeV which produces UTM magnitudes that are on average 50\% of the size of the missing transverse momentum in an event. Figure~\ref{fig:lowUTMmomentadistributions} shows the distributions of visible daughter transverse momenta, missing transverse momenta and event UTM, for a typical 10,000 event ``low-UTM'' run.

The algorithm was tested against many hundreds of thousands of signal-type Monte Carlo generated events of the form described above, and has shown good agreement with the best numerical techniques available: the fraction of evaluations in which the algorithm was found to generate $\mttwo$ values differing by more than 0.01GeV (or by order 0.01\%) from the known correct values for the events generated (see later description) was found to be \textit{O}($10^{-5}$ -- $10^{-4}$).
\subsection*{Observations}

It is interesting to note that the root values (remember these are actually equal to $\sin\theta$ where $\theta$ is the angle required to rotate vectors $\bf p$ and $\bf q$ away from $\bf a$ and $\bf b$ until equations \eqref{eq:parameterisedbalanced} and \eqref{eq:parameterisedshortsplitting} are satisfied) are very skewed towards a value for $\theta$ of 180 degrees. Figure~\ref{fig:lowUTMcosthetadistrib} shows the ``low UTM'' distribution for $\cosTheta$ (cosine of the rotation angle is used for clarity) demonstrating that an angle of approximately 180 degrees (i.e. $\cosTheta  = -1$) is highly favoured.   


To understand why $\sim180$ degrees is the dominant solution to the $\mttwo$ quartic expression in the fully massless case we need to look again at the geometric picture of what is happening with respect to momentum vectors $\bf a$, $\bf b$ and $\vecPtmiss$ (i.e.~${\bf p} +{ \bf q}$) depending on the exact size and direction of the UTM of the event.

\begin{figure}
\includegraphics[width=0.49\columnwidth]{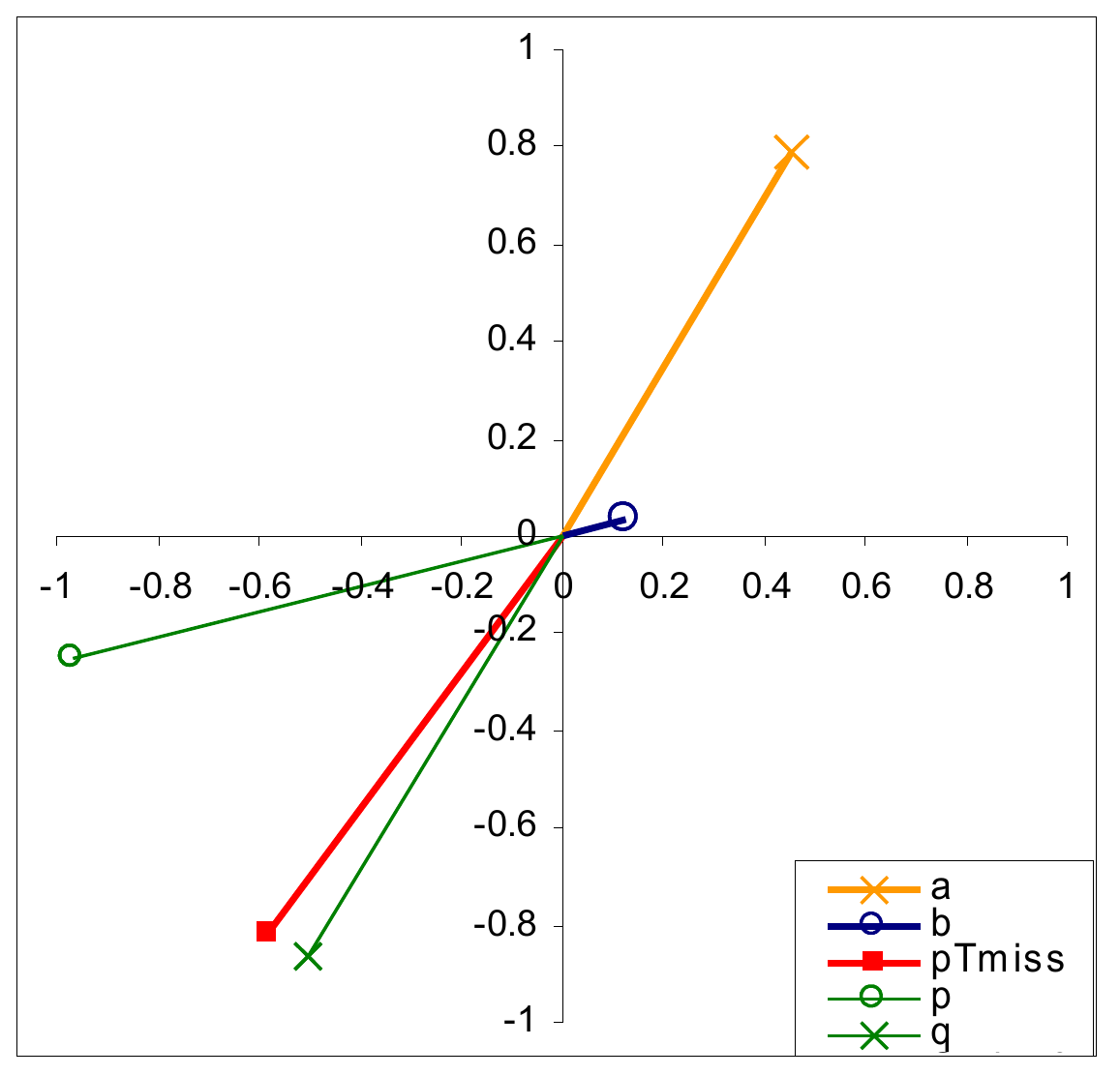} 
\includegraphics[width=0.49\columnwidth]{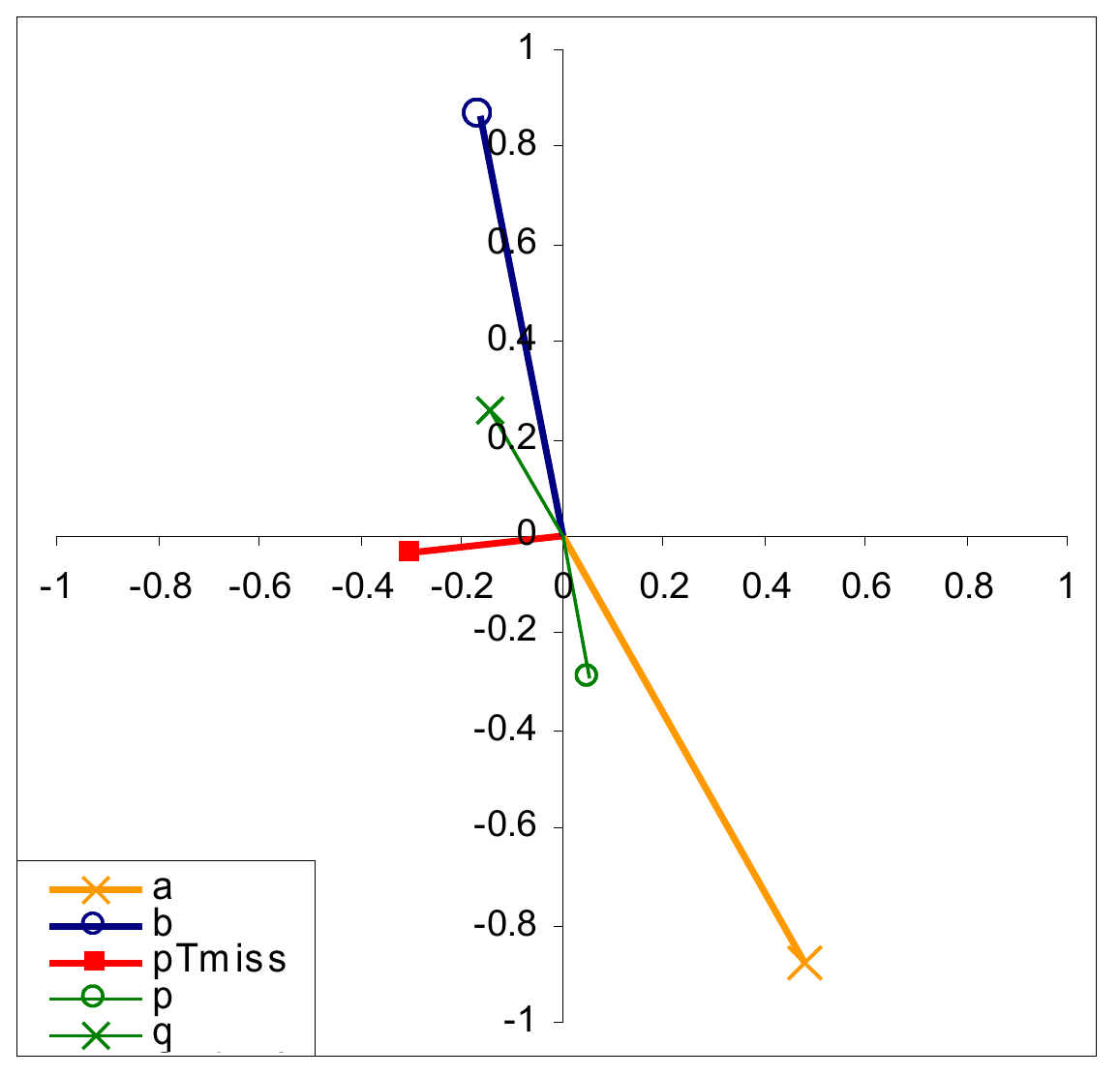} 
\caption{
Two sample events showing the transverse momenta unit-vectors where UTM is negligible. The left-hand plot shows an event where $\vecPtmiss$ is relatively large compared with the visible momenta {\bf a} and {\bf b} (thus {\bf a} and {\bf b} are ``pushed close together''). The right-hand plot shows an event where conversely $\vecPtmiss$ is relatively small versus $\bf a$ and $\bf b$ (thus $\bf a$ and $\bf b$ are more ``back-to-back''). The unit-vectors for the invisible momenta $\bf p$ and $\bf q$, determined by evaluating the quartic expression for the events, are also shown. Both of these events have a value of 180 degrees for the rotation angle $\theta$.
\label{fig:zeroUTMexampleunitvectors}}
\end{figure}

\subsection{An Important Result when the UTM = 0}

An important configuration occurs in this massless case when the UTM is zero (or negligible).  Equation~\eqref{eq:longsplitting} becomes
\begin{equation}
	{\bf a}+{\bf b}+\vecPtmiss={\bf 0}  \label{eq:longsplittingatzeroUTMa}
\end{equation}
hence
\begin{equation}
	{\bf a}+{\bf b}=-{\bf p}-{\bf q}  \label{eq:longsplittingatzeroUTMb}
\end{equation}
In order to continue to satisfy the requirement that the angle between \textit{q} and \textit{p} for the $\mttwo$ solution is fixed to be the same angular separation between \textit{a} and \textit{b} (and given the parameterisation of \eqref{eq:parameterq}) we have in this situation that

\begin{equation}	
	|{\bf a}| = |{\bf q}| \textup{ and } |{\bf b}| = |{\bf p}|	\label{eq:momconservedeq}
\end{equation}

\textit{and} that $\theta$ is exactly 180 degrees i.e. in the fully massless case, when UTM is zero, in order to conserve momentum, the parent particles decay such that the visible daughter of one ``side'' of the event has an equal but opposite momentum to the invisible daughter from the \textit{other} ``side'' of the event.  This means that the value of $\mttwo$ in the massless case when UTM is negligible, given by either side of expression \eqref{eq:parameterisedbalanced}, simplifies to

\begin{equation}
2\left|{\bf a}\right|\left|{\bf b}\right|\left(1-\cosTheta \ahdotbh - \sinTheta \eahbh\right)
\end{equation}
\begin{equation}
\label{eq:parameterisedbalancedzeroutm}
= 2\left|{\bf a}\right|\left|{\bf b}\right|\left(1+ \ahdotbh\right)
\end{equation}
hence
\begin{equation}
\mttwo^2 = 2 |{\bf a}||{\bf b}| (1 + \cos \phi) \label{eq:contransversemass}
\end{equation}
where $\phi$ is the angle between vectors $\bf a$ and $\bf b$. The expression on the right-hand side of this equation is actually also that for the so-called contransverse mass $\mct^2$ as defined in \cite{Tovey:2008ui}\footnote{This equality between $\mttwo$ and $\mct$ in the fully massless case when the UTM is zero has already been noted by Serna \cite{Serna:2008zk} using a more general derivation.}.

Figure~\ref{fig:zeroUTMexampleunitvectors} shows some sample Monte Carlo events where the UTM is negligible. It is clear visually why conservation of momentum leads to the relationships of \eqref{eq:longsplittingatzeroUTMb} and \eqref{eq:momconservedeq} between $\bf a$, $\bf b$, $\bf p$ and $\bf q$. 

\subsection{When the UTM $>$ 0}

The interesting result presented here showing that a majority of events tend to have $\theta$ close to 180 degrees is due to the fact that even when the UTM is non-zero, it will not necessarily change the rotation angle significantly.  This is because it is not just the magnitude, but also the specific direction of the UTM w.r.t. the directions of the visible and invisible momenta that determine if the UTM will have any affect on $\theta$. Specifically, the UTM would need to be both sufficiently non-parallel to one (or more) of the daughter particle momenta, but also of the order of or larger than that specific momentum vector's magnitude. Visually this can be thought of as the UTM ``pushing'' a smaller momentum vector to one side.

\begin{figure}
\includegraphics[width=0.49\columnwidth]{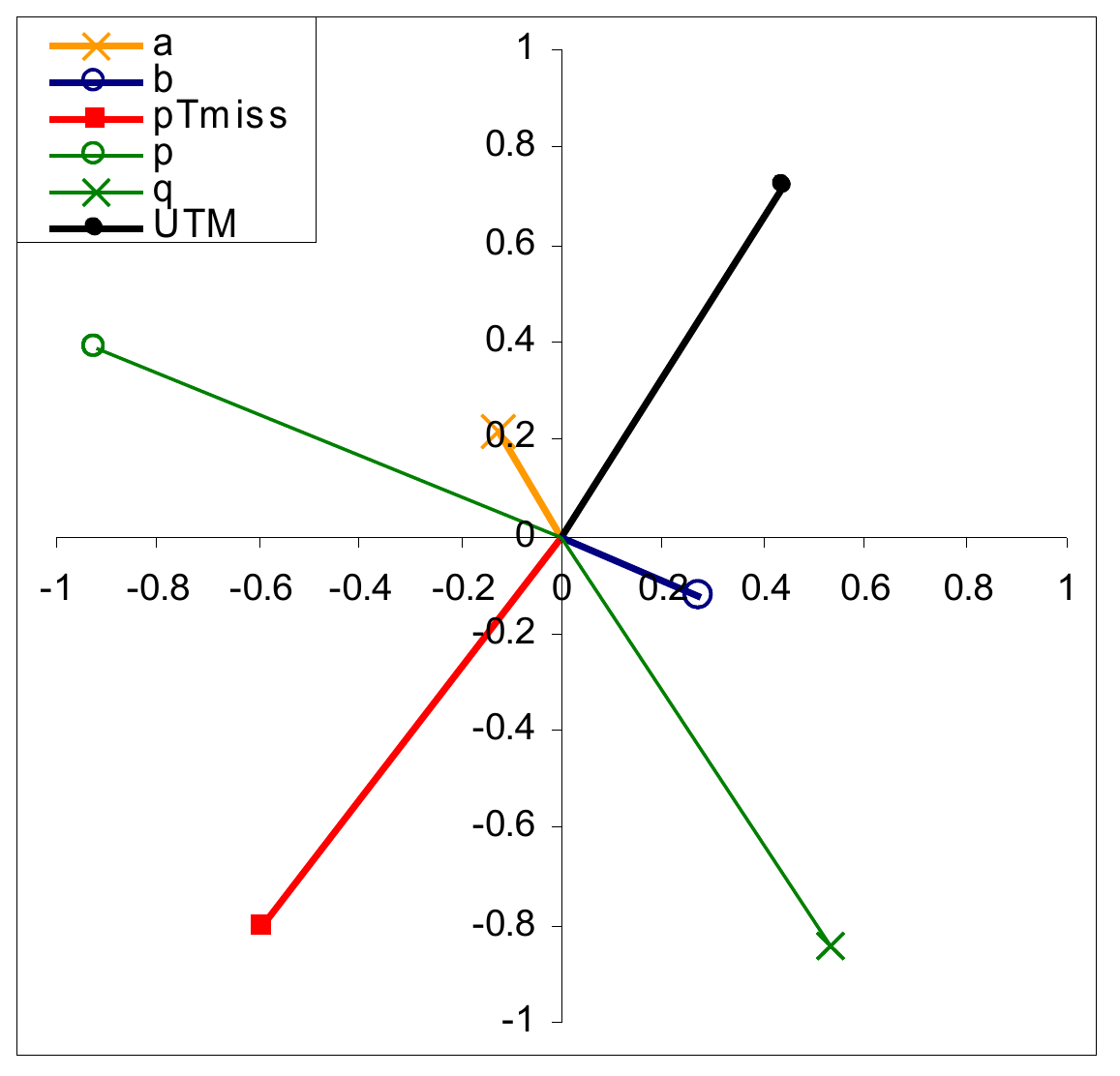} 
\includegraphics[width=0.49\columnwidth]{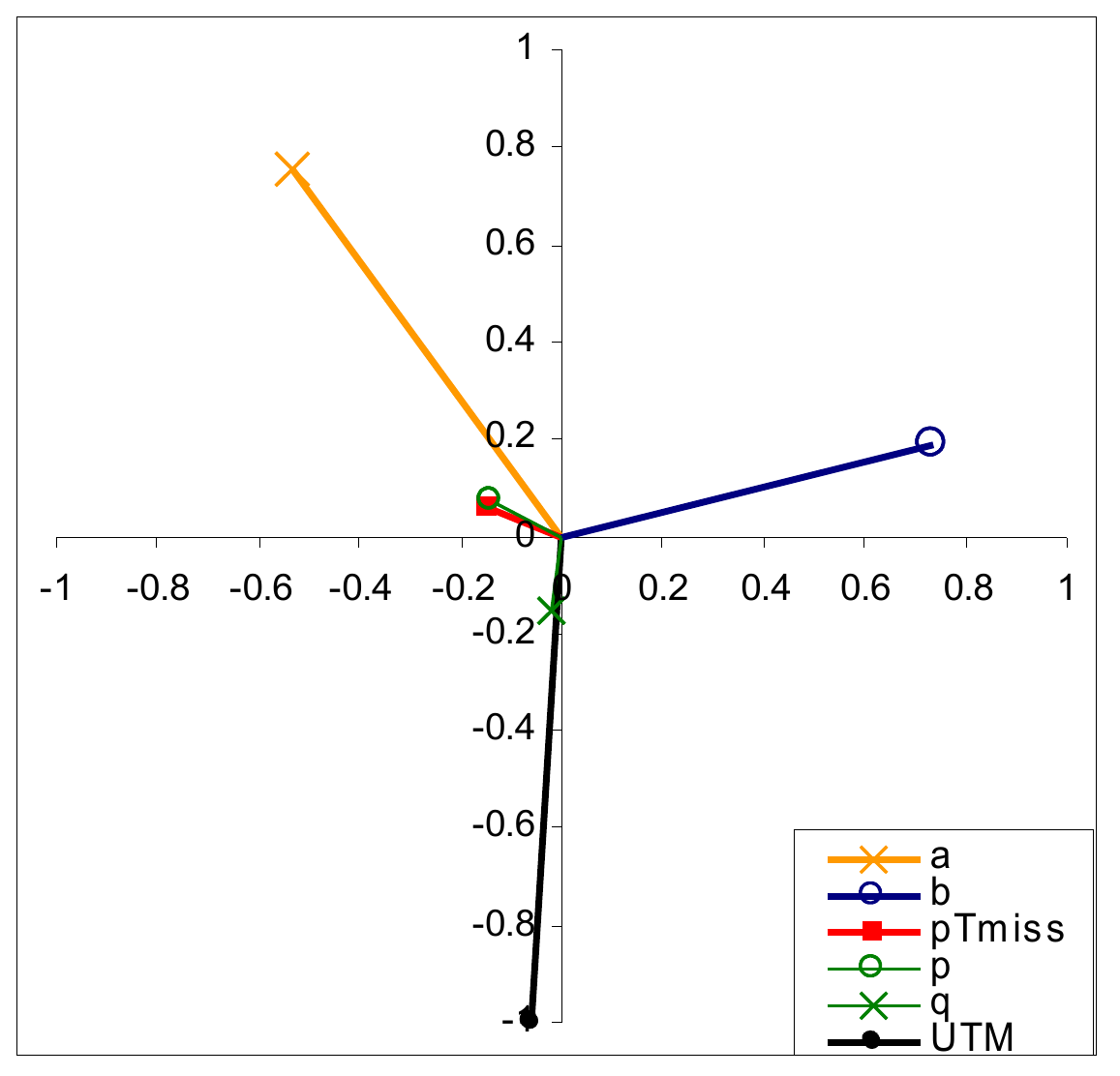} 
\caption{
Two situations where large UTM has (left plot) little effect, and (right plot) a bigger effect, on the rotation angle leading to the $\mttwo$ solution.
\label{fig:utmeffect}}
\end{figure}

For example, in figure~\ref{fig:utmeffect}, the left-hand plot is very similar to the left-hand plot in figure~\ref{fig:zeroUTMexampleunitvectors} (i.e. large $\vecPtmiss$ and relatively small visible momenta), except that here the UTM is significant (160~GeV versus 190~GeV for $\vecPtmiss$, 48~GeV for ${\bf a}$ and 58~GeV for ${\bf b}$). But because the UTM is roughly anti-parallel with $\vecPtmiss$, it has virtually no effect on the \textit{direction} of the missing transverse momentum and a minor effect on the angular displacements of {\bf a} and {\bf b} (actually the visible momenta will have been ``pushed away'' from each other somewhat compared with if there had been negligible UTM), and thus the appropriate rotation for ${\bf p}$ and ${\bf q}$ remains at close to 180 degrees. For this type of event to have a rotation angle significantly different from 180 degrees, the UTM vector would have needed to be at a substantial angle to the other momenta vectors i.e. in terms of momentum phase-space, only a relatively moderate sub-set of vector directions of UTM would be able to achieve this. The right-hand plot of figure~\ref{fig:utmeffect} shows an event of this type where the UTM has ``pushed'' vector {\bf b} (and also in this case the missing transverse momentum) away from it, thus decreasing the rotation angle for ${\bf p}$ and ${\bf q}$ to around 138 degrees.

\begin{figure}
\includegraphics[width=0.49\columnwidth]{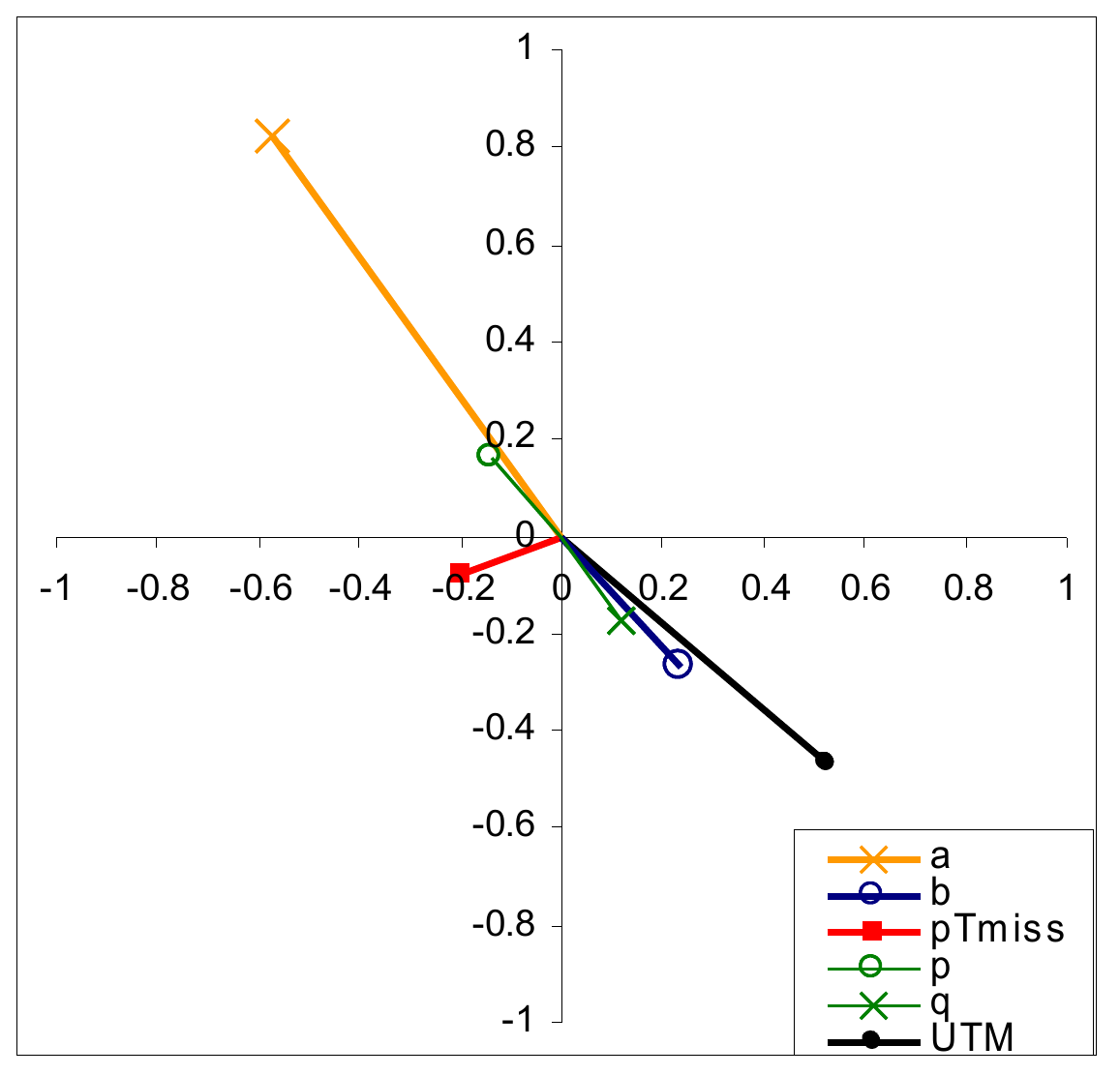} 
\includegraphics[width=0.49\columnwidth]{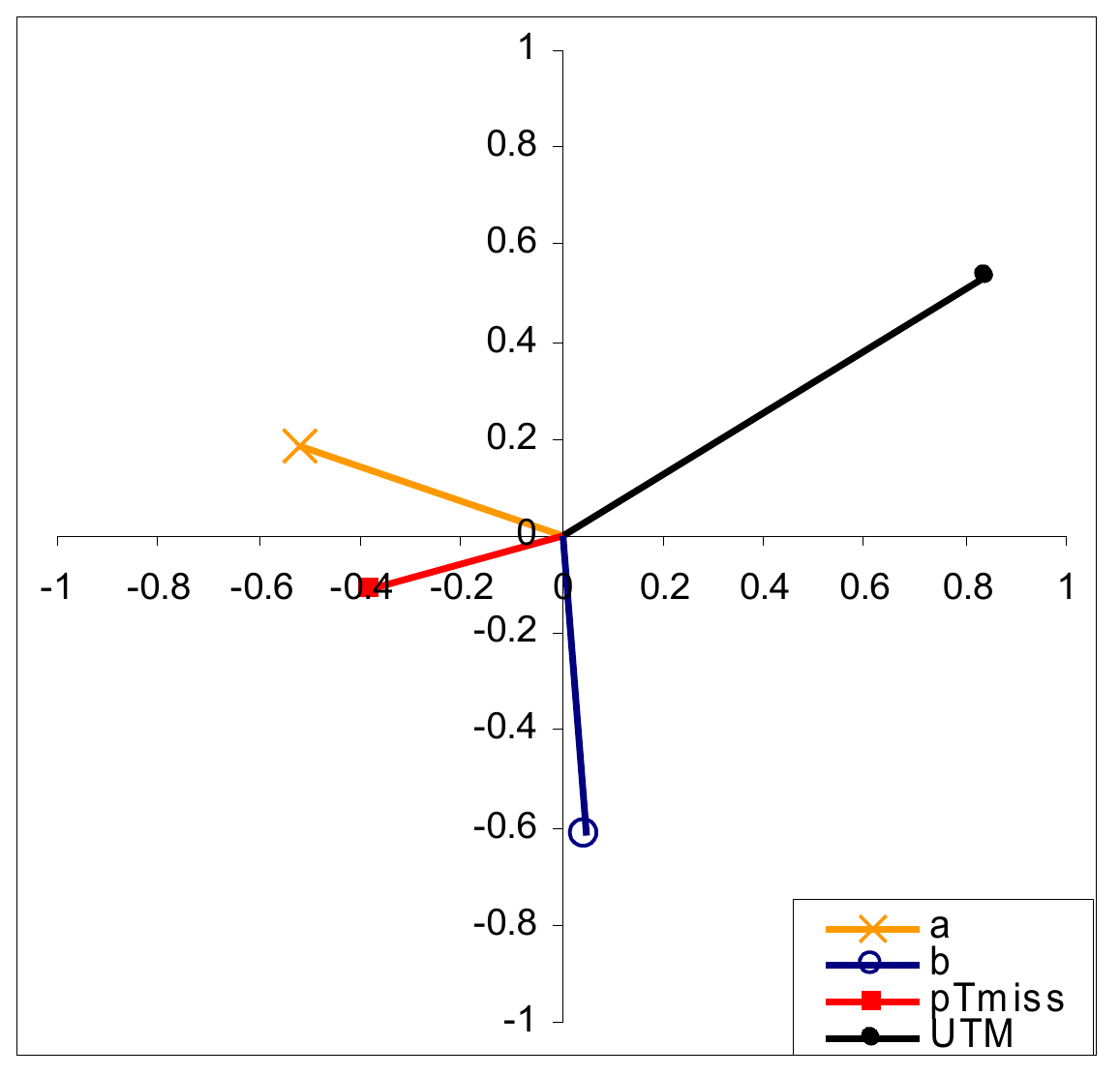} 
\caption{Small $\vecPtmiss$ events where a large UTM has (left plot) little effect on the rotation angle leading to the $\mttwo$ solution, and (right plot) a bigger effect but where the value of $\mttwo$ is now zero for each ``side'' of the event as $\vecPtmiss$ lies ``inside'' of ${\bf a}$ and $\bf b$.
\label{fig:backtobackutm}}
\end{figure}

One final comment to make on how UTM may or may not affect the rotation angle is to look at situations similar to the right-hand plot in figure~\ref{fig:zeroUTMexampleunitvectors} (i.e. where $\vecPtmiss$ is relatively small versus $\bf a$ and $\bf b$ and thus $\bf a$ and $\bf b$ are roughly ``back-to-back''). The left-hand plot of figure~\ref{fig:backtobackutm} shows again that a significant UTM is not on its own sufficient to move the rotation angle away from 180 degrees. Here a UTM of magnitude 71~GeV is roughly (anti-) parallel to the back-to-back $\bf a$ and $\bf b$ vectors (101~GeV and 36~GeV respectively), has therefore virtually no effect on their directions, and thus has no affect on the resulting rotation angle which continues to be approximately 180 degrees. 

However in this type of back-to-back event, even if the UTM has the right direction to ``push'' $\bf a$, $\bf b$ or $\vecPtmiss$ towards each other, if this were to cause $\vecPtmiss$ to now lie between $\bf a$ and $\bf b$, then we are back to the trivial zero situation where the balanced value for $\mttwo$ is identically zero on both sides of the event as mentioned in section~\ref{sec:findingroots} (the right-hand plot of figure~\ref{fig:backtobackutm} shows this type of event). So in this low $\vecPtmiss$, back-to-back situation, there is an even more limited phase-space for the UTM vector to move within in order to produce a rotation angle significantly different from 180 degrees.\footnote{Note therefore that in situations of higher UTM, we would expect there to be more trivial zero type events (i.e. $\mttwo$ values of zero).}

\section{Approximating $\mttwo$ using a first order correction term to $\theta$}
\label{sec:approximatingmt2}

Having understood that in the ``fully massless'' case there is a strong bias towards the rotation angle $\theta$ being close to 180 degrees, it is sensible to consider if the angle which leads to the correct balanced value for $\mttwo$ could be approximated by assuming $\theta$ is equal to 180 degrees plus a first order correction term; this approximation might possibly circumvent the need to solve the quartic equation at all.  

To investigate this we need to look again at the constraint equations \eqref{eq:parameterisedbalanced} and \eqref{eq:parameterisedshortsplitting} whose quotient eliminated $|{\bf p}|$ and $|{\bf q}|$ and which lead to the quartic expression. Looking at \eqref{eq:parameterisedbalanced} when $\theta$ is 180 degrees (remembering that this signifies an event where UTM is negligible) it is clear that it is the vanishing of the $\sinTheta$ term that allows the expression to be simplified (and led us to the derivation of the $\mct$ equivalent expression of \eqref{eq:contransversemass}).  When $\theta$ is not 180 degrees, it will therefore be this, now non-zero, $\sinTheta$ term which, as it has the opposite sign on either side of \eqref{eq:parameterisedbalanced}, will have the greatest effect on adjusting \eqref{eq:parameterisedbalanced} (and also \eqref{eq:parameterisedshortsplitting}) to allow the correct balanced values of $\mttwo$ to be calculated and the equality to continue to be satisfied.  

As an approximation, we will therefore assume that the rotation angle is represented by ($\theta$+$\delta\theta$), where $\delta\theta$ is a first order correction term for the angle. In what follows we will refer to this estimated rotation angle as $\theta_e$ (and to the true rotation angle, i.e. as determined from the quartic roots, as $\theta_t$) and we assume as a starting point that $\theta_e$ equals 180 degrees.  As $\delta\theta$ is by definition a small angular adjustment, then to a good approximation the cos($\theta+\delta\theta$) terms can be replaced by -1 and the sin($\theta+\delta\theta$) terms can be replaced with -$\delta\theta$. The two constraint equations would therefore simplify to:
\begin{equation}
\left|{\bf a}\right|\left|{\bf p}\right|\left(1+\ahdotbh + \delta\theta \eahbh\right)
=\label{eq:parameterisedbalancedsimple}
\left|{\bf b}\right|\left|{\bf q}\right|\left(1+\ahdotbh - \delta\theta \eahbh\right) 
\end{equation}

\begin{equation}
+\left|{\bf p}\right|\left(-\ebp -\delta\theta\bhdotph\right)
= \label{eq:parameterisedshortsplittingsimple}
-\left|{\bf q}\right|\left(-\eap -\delta\theta\ahdotph\right)
\end{equation}

Taking the quotient of both sides (again to eliminate any dependence on $|{\bf p}|$ and $|{\bf q}|$) and rearranging we obtain a quadratic function in $\delta\theta$:
\begin{equation}
	B(|{\bf b}|D - |{\bf a}|F )\delta\theta^2 + (|{\bf a}|(BE - AF) - |{\bf b}|(AD + BC))\delta\theta + A(|{\bf a}|E + |{\bf b}|C) = 0
\end{equation}
in which
\begin{eqnarray}
	A &=& (1 + \ahdotbh) \\
	B &=& \eahbh \\
	C &=& -\ebp \\
	D &=& \bhdotph \\
	E &=& -\eap \\
	F &=& \ahdotph	
\end{eqnarray}

This quadratic equation can be solved using the usual quadratic formula\footnote{The sign of the $\delta\theta$ term is assumed to signify which of the two roots is the correct physical value (i.e. whether to choose the $+$ or $-$ in the quadratic formula).} and the value for $\delta\theta$ thus gives a correction to our assumed 180 degree value for $\theta_e$. This corrected value can then be used to calculate approximate balanced values for $\mttwo$. We will refer to these approximated values going forward as \thenewvar\ values.  We would expect in the cases where $\theta_t$ is close to 180 degrees, that our approximated value for $\theta_e$, and thus \thenewvar, would be very close to the ``true'' values calculated using the quartic algorithm.  In the rarer events where the UTM produces a rotation angle significantly different from 180 degrees (say more than 45 degrees different), we would not expect the correction term to produce a particularly accurate estimation of $\theta_t$ and thus $\mttwo$. However in these events, we can again use the geometry of the massless case to estimate the correct rotation angle:

\mbox{}

As discussed in section~\ref{sec:analysis}, it is the symmetric distribution of the visible and invisible momenta in low UTM events which leads to rotation angles close to 180 degrees. Conversely because events with $\theta_t$ significantly different from 180 degrees is caused by a UTM vector which ``pushes'' the visible and invisible momenta closer together, these events are by definition asymmetric and tend to have ${\bf a}$ and ${\bf b}$ vectors ``close together'' (the right hand plot of figure~\ref{fig:utmeffect} shows this typical situation).  We can make a simple estimation of the correct rotation angle by choosing one of three possible scenarios (i) a rotation angle that is just more than the minimum possible angle (i.e.~that which ensures that ${\vecPtmiss}$ is just inside ${\bf p}$ and ${\bf q}$), (ii) a rotation angle which is just less than the maximum possible angle (again ensuring ${\vecPtmiss}$ remains inside ${\bf p}$ and ${\bf q}$), or (iii) a rotation angle that sits ${\vecPtmiss}$ halfway between ${\bf p}$ and ${\bf q}$. To determine which is most appropriate we can, for example, look at the relative magnitudes of ${\bf |a|}$ and ${\bf |b|}$, as for the balanced situation (i.e. see equation~\eqref{eq:balancedeq}) we would expect a dominant $\hat {\bf a}$ vector (say where ${\bf |a|}$ is more than twice the size of ${\bf |b|}$) to be associated with a dominant $\hat {\bf q}$ (i.e. from the parametrisation of $\hat {\bf q}$ in \eqref{eq:parameterq}) and a dominant $\hat {\bf b}$ to result in a dominant $\hat {\bf p}$ (from equation~\eqref{eq:parameterp}); the right hand plot of figure~\ref{fig:utmeffect} shows an example of this latter scenario. By looking at the relative magnitudes of ${\bf |a|}$ and ${\bf |b|}$ for an event, we can therefore choose the minimum or maximum angle that ensures the appropriate vector dominates ${\vecPtmiss}$, and thus estimate $\mttwo$ for that event. If neither visible momentum vector is dominant, we choose scenario (iii) which estimates equal ${\bf |p|}$ and ${\bf |q|}$ values.

\begin{figure}
\mbox{\subfigure[ Momenta distributions in the ``Mid UTM'' scenario]{\includegraphics[width=0.49\columnwidth]{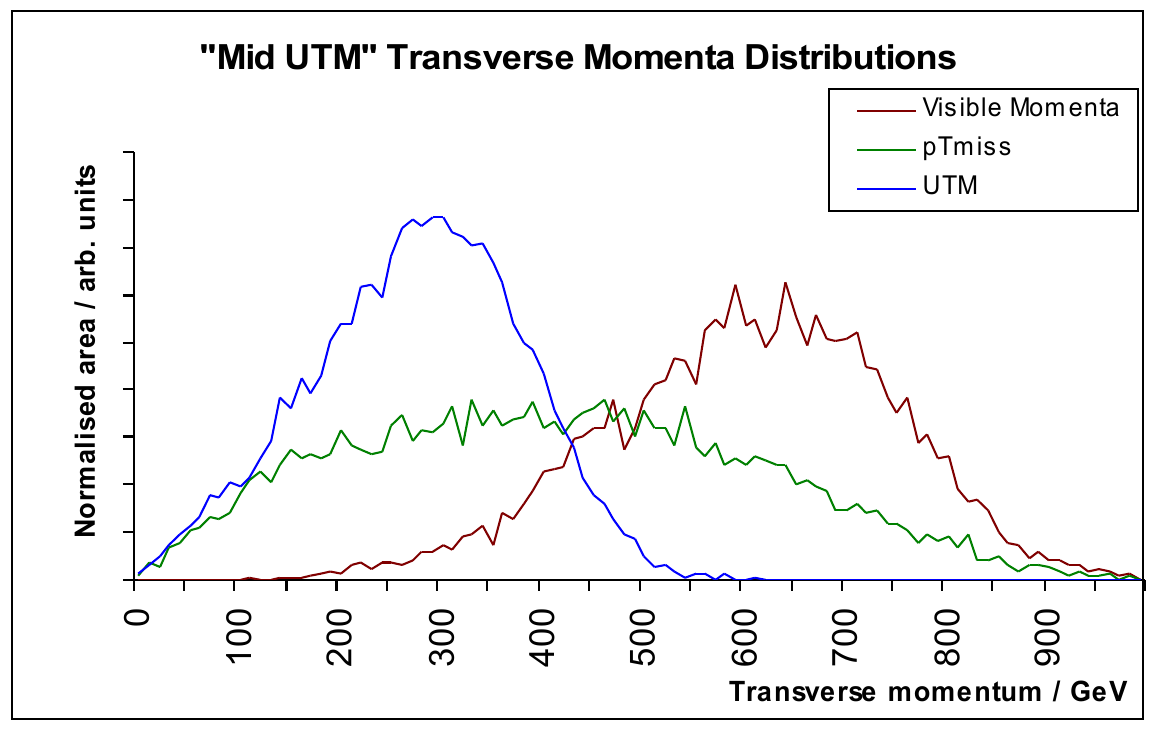}
\label{fig:midUTMmomentadistributions}}
\subfigure[ Momenta distributions in the ``High UTM'' scenario]{\includegraphics[width=0.49\columnwidth]{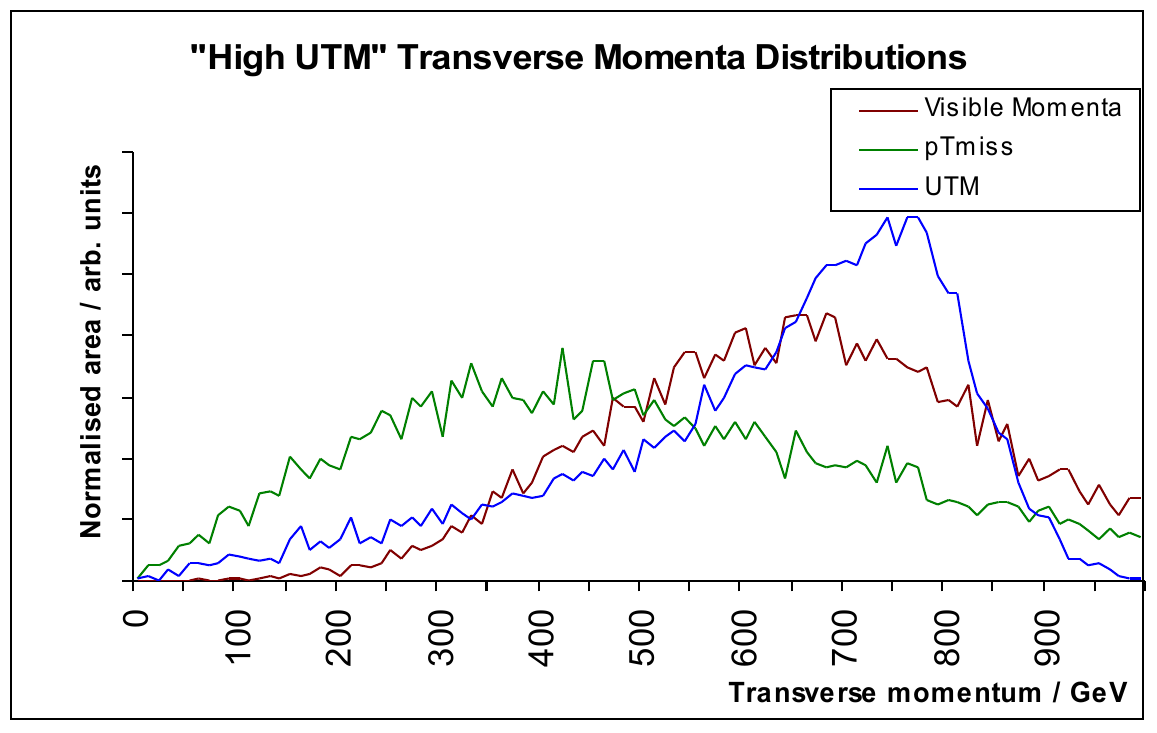}
\label{fig:highUTMmomentadistributions}}}
\caption{
Momenta distributions of higher UTM scenarios
\label{fig:midhighmomentadistributions}}
\end{figure}

\begin{figure}
\mbox{\subfigure[ $\cosTheta_t$ distributions in the ``Mid UTM'' scenario]{\includegraphics[width=0.49\columnwidth]{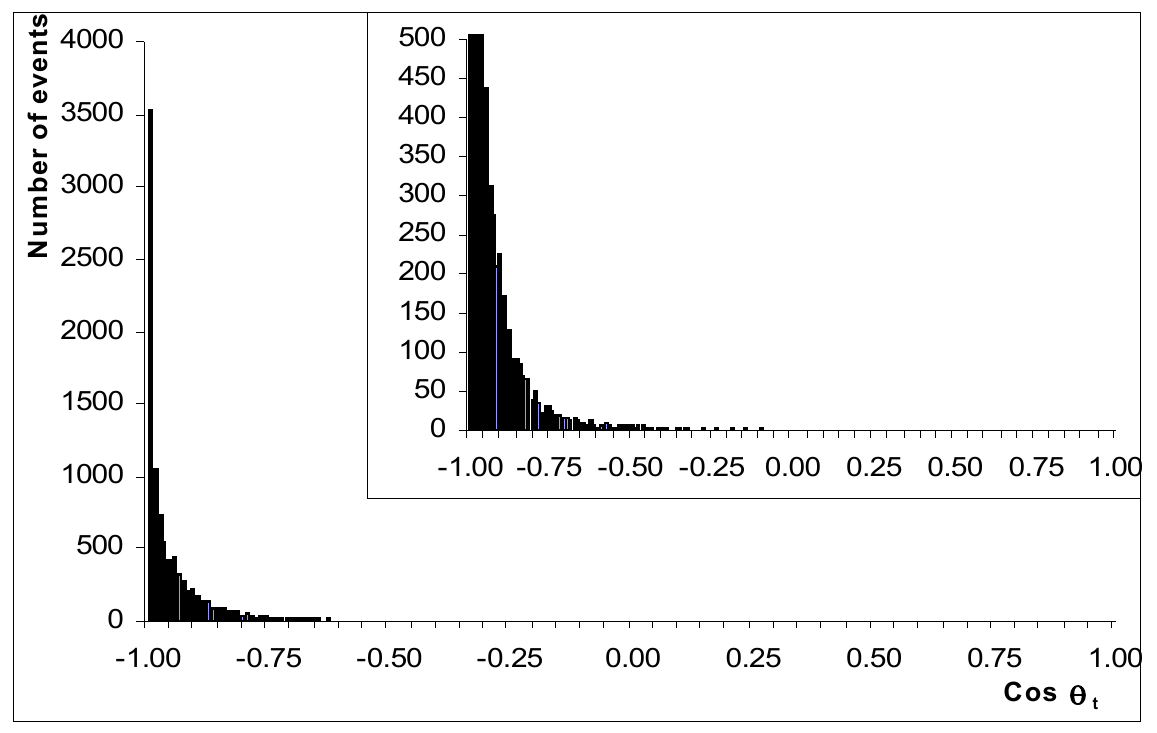}
\label{fig:midcosthetdistributions}}
\subfigure[ $\cosTheta_t$ distributions in the ``High UTM'' scenario]{\includegraphics[width=0.49\columnwidth]{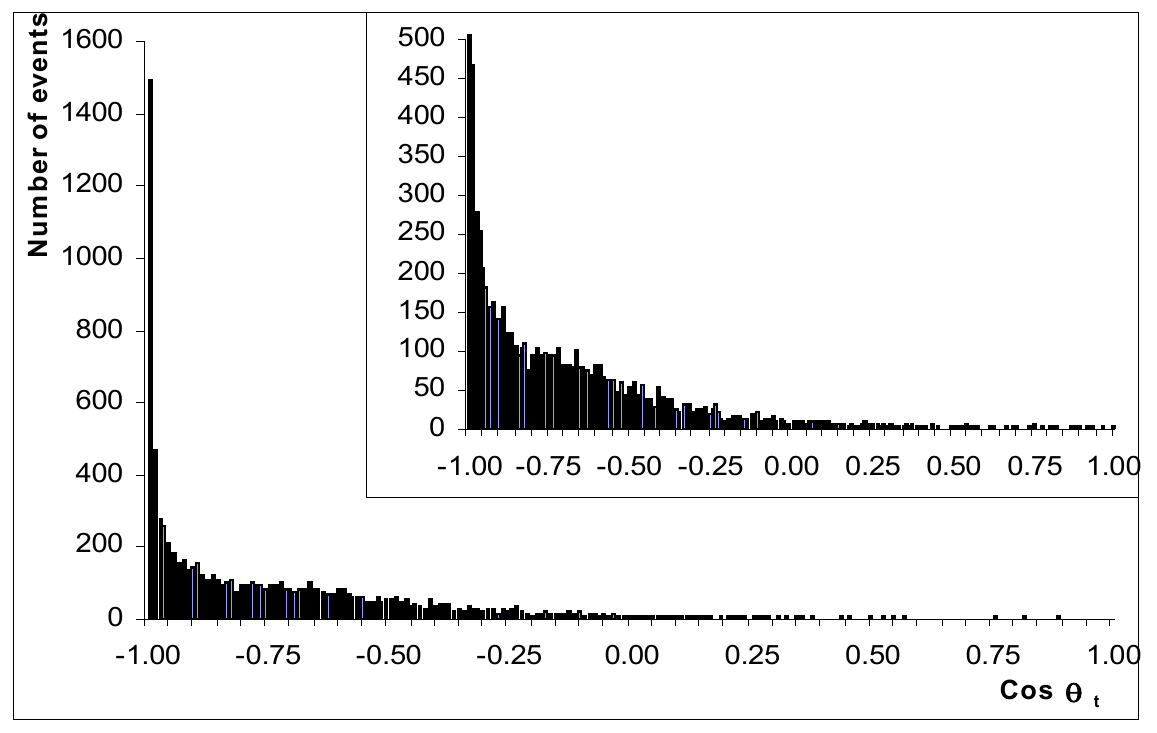}
\label{fig:highUTMcosthetadistributions}}}
\caption{
$\cosTheta_t$ distributions of higher UTM scenarios
\label{fig:midhighcosthetadistributions}}
\end{figure}

By adding this estimation method to the \thenewvar\ approximation, we now have an algorithm which we hope can provide a fast calculator for estimating $\mttwo$ irrespective of the rotation angle in question.  In order to check the algorithm's accuracy, in addition to testing it in the ``low UTM'' scenario discussed previously, the same toy Monte Carlo generator was used to produce events in two other scenarios: the ``mid UTM'' run where the UTM in an event is on average equal to the typical $\vecPtmiss$ magnitude, and the ``high UTM'' run, where the UTM is on average twice the size of the typical $\vecPtmiss$ magnitude.  The distributions for visible daughter transverse momenta, missing transverse momenta and UTM for these higher UTM scenarios are shown in figure~\ref{fig:midhighmomentadistributions}.  The resulting values of $\cosTheta_t$ obtained for these two runs using the quartic calculator are shown in figure~\ref{fig:midhighcosthetadistributions}. The longer tails should be compared with the previous figure~\ref{fig:lowUTMcosthetadistrib} demonstrating that scenarios of higher UTM are more likely to produce rotation angles significantly different from 180 degrees. However if should also be noted that even in these higher UTM regimes, the rotation angle distribution is still dominated by events with $\theta_t$ close to 180 degrees as expected and as discussed in section~\ref{sec:analysis}.

\begin{figure}
\includegraphics[width=0.49\columnwidth]{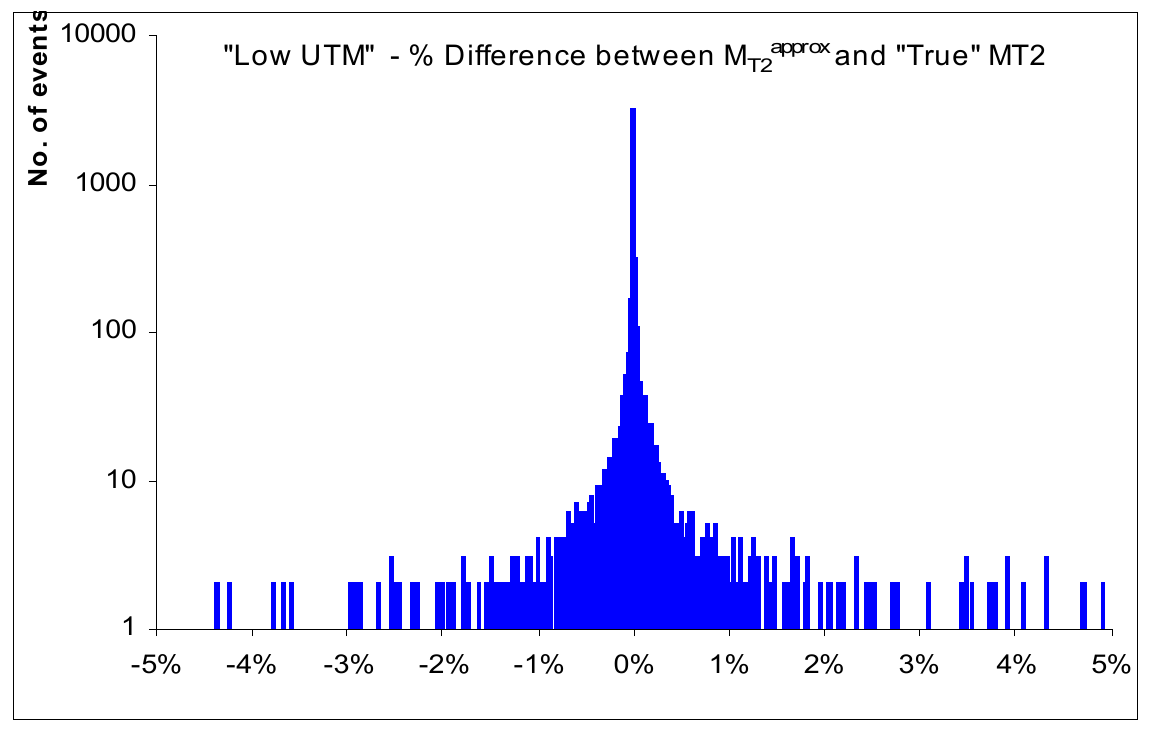}
\caption{
Percentage difference between \thenewvar\ and $\mttwo$ in the ``low UTM'' scenario 
\label{fig:comparitivedistrosofLowUTMdeltas}}
\end{figure}

The approximate method was used to calculate \thenewvar\ for 10,000 ``mid UTM'' and ``high UTM'' events as well as the original 10,000 lower UTM events, and these were compared with the ``true'' values as obtained with the quartic calculator. Firstly Figure~\ref{fig:comparitivedistrosofLowUTMdeltas} shows the resulting distribution of the percentage difference between the ``low UTM'' \thenewvar\ value and the ``true'' $\mttwo$ value for each event (percentage difference calculated by subtracting the ``true'' $\mttwo$ value from the \thenewvar\ value and dividing by the ``true'' $\mttwo$ value - number of events per bin plotted on a log scale to see the distribution tail more clearly).  The accuracy of the \thenewvar\ algorithm is very encouraging with over 95\% of events within 2\% of the ``true'' $\mttwo$ value.

\begin{figure}
\mbox{\subfigure[ Percentage difference between \thenewvar\ and $\mttwo$ in the ``Mid UTM'' scenario]{\includegraphics[width=0.49\columnwidth]{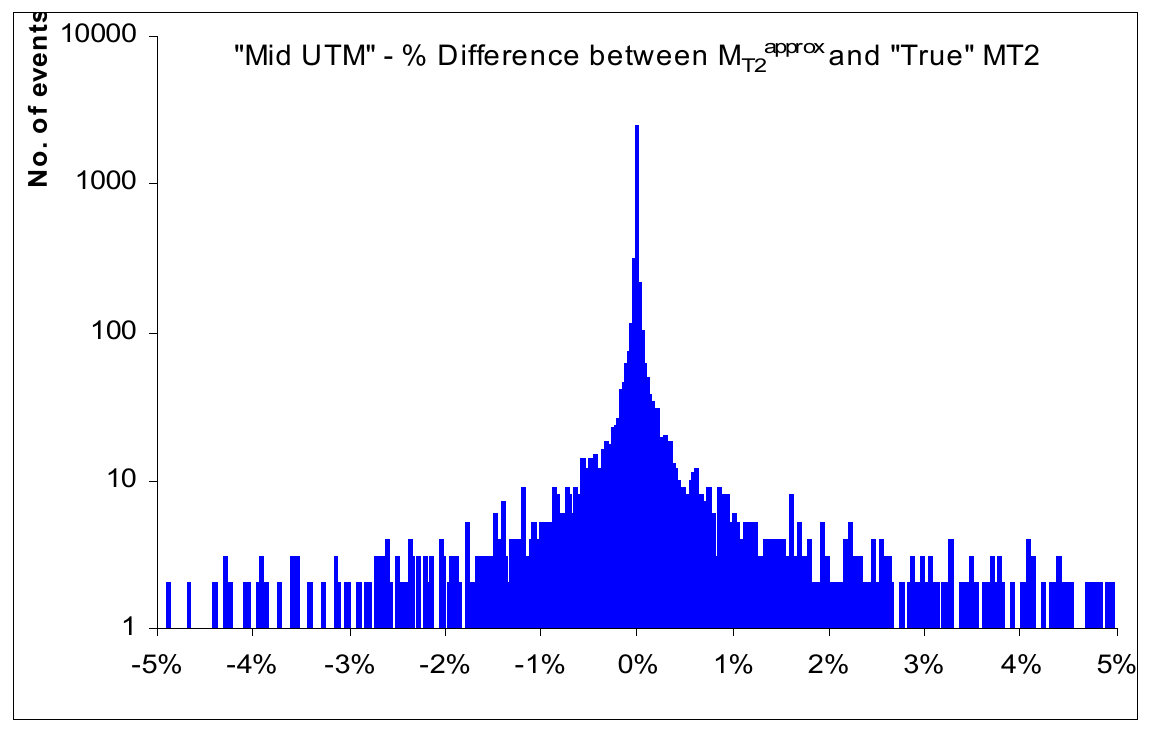}
\label{fig:midUTMSimpleMt2comparison}}
\subfigure[ Percentage difference betweem \thenewvar\ and $\mttwo$ in the ``High UTM'' scenario]{\includegraphics[width=0.49\columnwidth]{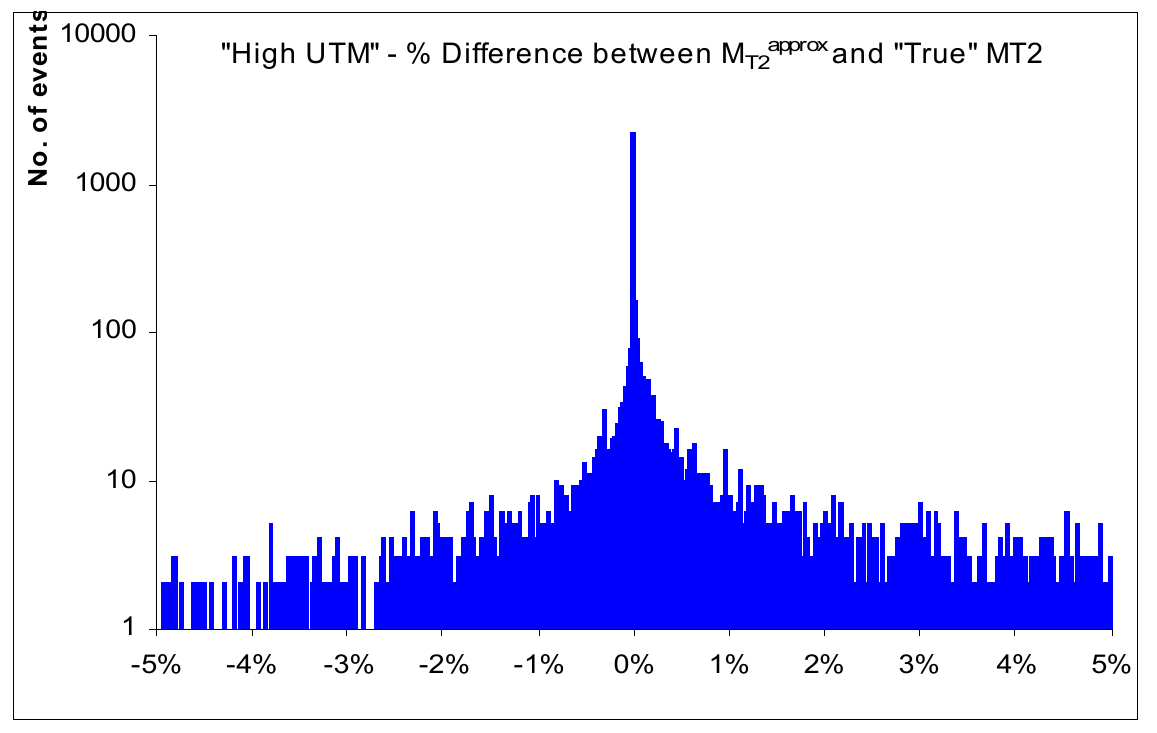}
\label{fig:highUTMSimpleMt2comparison}}}
\caption{
Percentage difference between \thenewvar\ and ``true'' $\mttwo$ in the higher UTM scenarios
\label{fig:comparitivedistrosofHighUTMdeltas}}
\end{figure}

\pagebreak
Figure~\ref{fig:comparitivedistrosofHighUTMdeltas} shows the same comparisons for the two higher UTM runs\footnote{In the ``high UTM'' run, it is not surprising that approximately 16\% (1,626) were of the ``trivial zero'' type with $\mttwo = 0$ (compared with less than 10\% (837) in the ``mid UTM'' run). The histograms here therefore exclude these ``trivial zero'' events and show 8,374 events (high UTM) and 9,163 events (mid UTM).}.  As would be expected from the higher UTM in these events (leading to more events with rotation angles significantly different from 180 degrees), the \thenewvar\ values have larger tails than the ``low UTM'' situation but still demonstrate very good accuracy: for the ``mid UTM'' run about 90\% of values lie within 2\% of the ``true'' $\mttwo$ value, and for the ``high UTM'' run over 70\% of them are within 2\% of the ``true'' value (and 90\% within 10\% of the ``true'' value).

Although the \thenewvar\ calculator demonstrates good accuracy for estimating the $\mttwo$ value for an event, the real benefit it brings is one of speed. 

\section{Comparison of the speed of the ``true'' $\mttwo$ and \thenewvar\ calculators with the fastest numerical calculator}

For the ``true'' $\mttwo$ calculator, where $\mttwo$ is determined from the physical root of the underlying quartic equation, a typical calculation (where transverse momentum values ${\bf a}$, ${\bf b}$ and $\vecPtmiss$ are provided to the code) takes a little less than 3$\mu$s per event to run on a standard PC (ignoring any read/write times to and from datafiles).  This is approximately 4 times faster than the current fastest numerical $\mttwo$ calculating code provided by \cite{Cheng:2008hk} in their ellipse bisecting procedure (used in the massless case). One limitation on speed for the ``true'' $\mttwo$ algorithm seems to be the trigonometric nature of the quartic solving procedure requiring a number of library calls within the code. The authors are inclined to believe that more experienced programmers could modify the current code and improve this initial performance. 

For the \thenewvar\ algorithm, however, this analytic method can calculate a good estimate of $\mttwo$  in less than 1$\mu$s per event on a standard PC; that is more than 10 times faster than the current fastest numerical calculator.

\section{Use of $\mttwo$ in the fully massless case as a trigger variable}

As discussed in \cite{Lester:2011nj}, one of the motivations for finding useful ``special cases'' in which the solution for $\mttwo$ is fully analytic and can therefore be evaluated quickly and accurately, is its possible use as a ``trigger'' variable at the LHC.  The properties of $\mttwo$ as a variable which is sensitive to the mass-scale of the pair-produced parents and its ability to discriminate larger mass-scale events from low multiplicity, relatively low-mass-scale, QCD events, would indicate that $\mttwo$ should perform well as a trigger to help the LHC experiments cope with the ever-increasing instantaneous luminosity conditions.  

As has been mentioned, the new analytic calculator presented here is the fastest yet for calculating $\mttwo$ in the fully massless case (and improvements in speed may be possible still), and with regards to the \thenewvar\ approximation discussed in section~\ref{sec:approximatingmt2}, this has been shown to be a good approximation to the true $\mttwo$ value and, as a much simpler calculation, may be fast enough to compete with other possible trigger variables available. 

\begin{figure}
\mbox{\subfigure[ Difference between event Parent Mass and both \thenewvar\ and $\mct$ in the ``Mid UTM'' scenario]{\includegraphics[width=0.49\columnwidth]{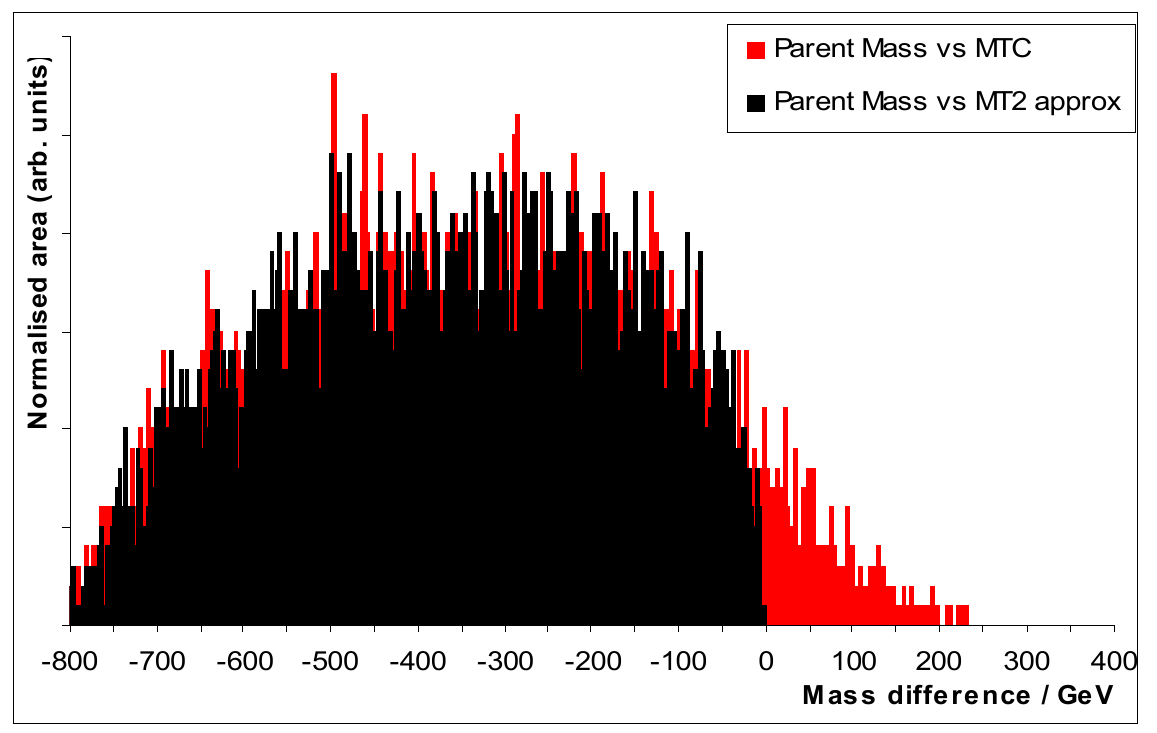}
\label{fig:midUTMSimpleMt2Masscomparison}} 
\subfigure[ Difference between event Parent Mass and both \thenewvar\ and $\mct$ in the ``High UTM'' scenario]{\includegraphics[width=0.49\columnwidth]{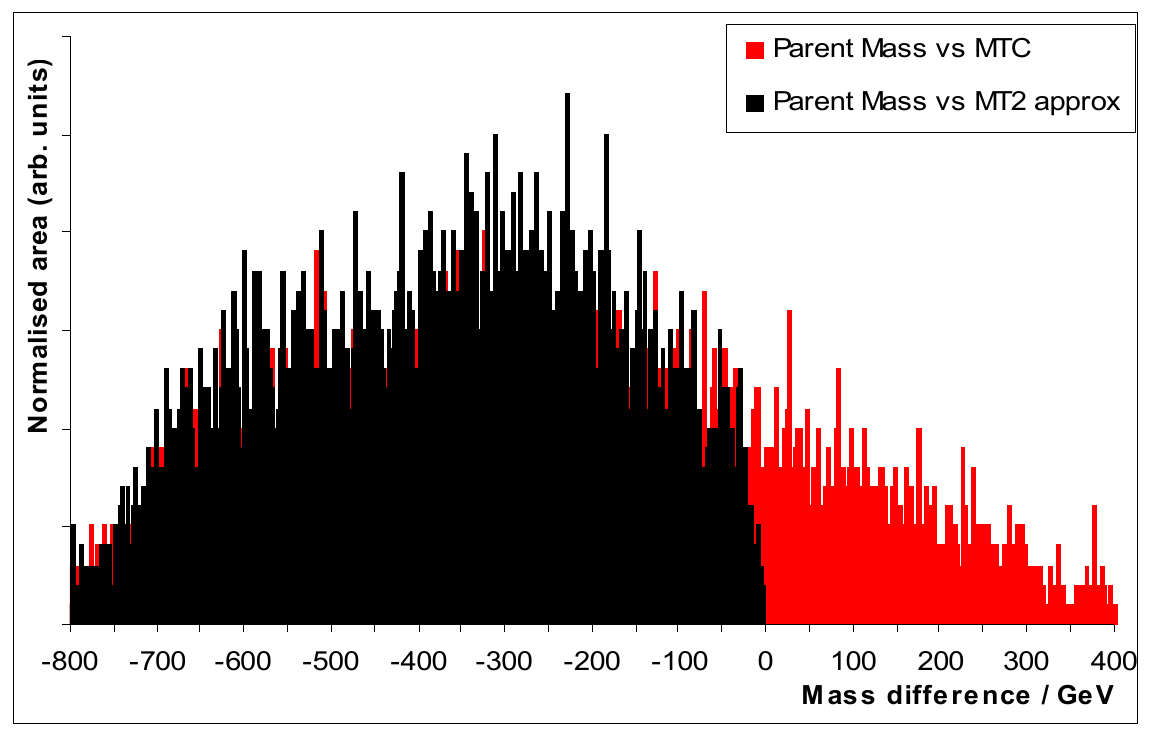}
\label{fig:highUTMSimpleMt2Masscomparison}}}
\caption{
Comparison of \thenewvar\ and $\mct$ as mass-scale identifiers
\label{fig:massdifferencemctmt2versusmassforhighandmidutm}}
\end{figure}

As an example of its sensitivity, we have compared its performance as a mass-scale indicator against that of $\mct$ (the contransverse mass, which in section~\ref{sec:analysis} we noted is equal to $\mttwo$ in the fully massless case when the UTM is exactly zero).  For the ``mid'' and ``high UTM'' runs, we compare the value of $\mct$ and \thenewvar\ against the rest mass of the pair-produced parent particle giving rise to the decay being analysed (remember this was chosen to be a normally distributed set of mass values with a mean of 750~GeV).  Figure~\ref{fig:massdifferencemctmt2versusmassforhighandmidutm} shows the mass difference between both $\mct$ and \thenewvar\ and the parent mass for both the ``mid UTM'' and ``high UTM'' runs. As would be expected from its construction as a maximal lower-bound, $\mttwo$ would never give a higher mass than that of the parent and it is pleasing to see that \thenewvar\ also demonstrates this quality, all events being less than or equal to zero mass difference. However $\mct$ has a significant tail demonstrating that, based as it is on the visible momenta of an event only, that when UTM is non-zero and thus the magnitudes of the visible momenta are larger than they otherwise would be to balance this UTM, $\mct$ tends to overestimate the mass-scale of an event.  

\section{Conclusions}

We have constructed a fast and stable analytic calculator for $\mttwo$ in the ``fully massless'' case (the scenario in which $\mttwo$ has been most frequently used at the LHC) based on the solving of a quartic equation.  Unfortunately due to the trigonometric nature of the quartic solution and the necessary library calls required by the code, the speed of this calculator is only around 4 times faster than the fastest available numerical code due to \cite{Cheng:2008hk} (though we will continue to work on the algorithm and believe that further improvements in speed are achievable). However due to the kinematics characteristic of the ``fully massless'' case we have also been able to approximate $\mttwo$ using a much faster and more straightforward algorithm that requires little more than the solving of a quadratic equation.  This approximation, \thenewvar, has been shown to be an accurate estimator of $\mttwo$, and \thenewvar\ values can be calculated more than 10 times faster than the current fastest numerical calculator.  

It was the authors' hope in beginning this analysis of $\mttwo$ in the ``fully massless'' case that it might lead to a $\mttwo$ calculator that could be fast enough to use as a ``trigger'' variable at the LHC and next linear collider.  The \thenewvar\ algorithm may be a possible step in that direction and it's usefullness as a mass-scale identifier has been demonstrated.

\section{Acknowledgments}
The authors would like to thank The Ogden Trust, Cranbrook School and the colleges of Corpus Christi and Peterhouse in the University of Cambridge for their support in allowing this work to be undertaken, and to RWD Nickalls for helpful discussions.

\section{Appendix : Quartic Root Finding \label{sec:app}}

\label{sec:solvingquartic}

Methods for solving quartics analytically\footnote{Quartics being the highest degree polynomial which \textit{can} be solved analytically.} have been known for centuries and Ferrari is credited with achieving it first in 1540 \cite{CardanoArsMagna}. The methods (all essentially based on Ferrari's original \cite{Shmakov:2011aa}) involve manipulation of the quartic coefficients in order to generate a so-called ``resolvent'' cubic. As with all polynomials, the roots of cubic and quartic equations can be real or complex. In fact the three roots of the resolvent cubic are either all real (leading either to four real quartic roots or to two pairs of complex conjugate quartic roots), or have one real and two complex conjugate roots (leading to two real and one pair of complex conjugate quartic roots).

Unfortunately, due in part to the number of algebraic steps required, algorithms for solving quartic equations tend to suffer from instability \cite{HerbisonEvans:2011aa}. These instabilities can arise due to, for example, numerical round-off errors in the calculations or the handling of imaginary numbers. Two situations relevant to this analysis are:

\begin{itemize}
\item
 where the resolvent cubic has three real \textit{positive} roots (and the quartic should therefore have four real roots) but rounding errors cause one or more of the smallest of these cubic roots to appear to be negative - this error will lead the algorithm to obtain incorrect \textit{complex} roots for the quartic.
\item
 where the resolvent cubic has one real and two complex conjugate roots but rounding errors cause the imaginary values of the complex conjugate pair to be unequal - this will lead to four complex roots for the quartic, when in fact two of the roots should be real.
\end{itemize}

As we know \textit{a priori} that the quartic expression for $\mttwo$ must have at least two real roots (i.e. one of the two real roots will lead to the physical value for $\mttwo$), we require a method that makes explicit (a) whether or not a given quartic has any complex roots - this enables us to identify immediately rounding errors such as in the first situation above\footnote{ i.e.\ if the quartic has \textit{all} real roots then we know the resolvent cubic must have three real \textit{positive} roots and we can proceed accordingly.}, and (b) the process by which the three resolvent cubic roots are combined to produce the four quartic roots - this allows us both to correct any rounding errors to the imaginary parts of the resolvent cubic's complex roots, and to minimise the need to manipulate the imaginary parts of any complex roots, as we as we know these must cancel when forming the two real quartic roots (i.e. we are only interested in any \textit{real} roots). 

A method developed by Leonard Euler in the eighteenth century \cite{Bell:2008le,EulerElements}, recently expanded upon by Nickalls \cite{Nickalls:2009aa},
allows the following:

\begin{itemize}
\item
 A simple way to calculate the discriminant of the quartic equation thus showing immediately if the roots of the quartic are all real or if there are two real/two complex roots (and thus also determining if the resolvent cubic's roots are all real or one is real and the others form a complex conjugate pair).
\item
A relationship between the resolvent cubic roots and those of the quartic is given explicitly thus allowing a stable algorithm to be constructed that derives the latter from the former without concern for rounding errors in, or any significant manipulation of, imaginary numbers.
\end{itemize}

Furthermore,  Nickalls \cite{Nickalls:1993aa} gives two complementary methods for calculating the roots of the resolvent cubic depending on whether or not there are one or three real roots. The benefit this brings to the algorithm is that in the situation where we know we have only real roots, we can use a more stable trigonometric calculation to obtain them.

The rest of this appendix summarises the approaches detailed in \cite{Nickalls:2009aa} and \cite{Nickalls:1993aa} and which underlie the resulting code for calculating the roots to the $\mttwo$ quartic expression.

\subsection{Resolvent cubic}
Leonard Euler was the first person to fully appreciate to role of the so-called ``resolvent'' cubic in calculating the roots of quartic equations \cite{Bell:2008le,EulerElements}.  Nickalls has written at length on the geometry of, and solutions to, various polynomials, and the details in this appendix are from two particular papers \cite{Nickalls:2009aa,Nickalls:1993aa}. The notation follows that in \cite{Nickalls:2009aa}, but sometimes also shows the notation in \cite{Nickalls:1993aa} for ease of reference\footnote{The rationale for, and derivations of, expressions which appear in this appendix can be found in the relevant references. This appendix will simply set out the procedure for finding the quartic roots.}.

For any given quartic of the form
\begin{equation}	
ax^4 + bx^3 + cx^2 + dx + e = 0	\label{eq:generalquartic}
\end{equation}

manipulation of the coefficients of the quartic allows the ``reduced'' form of the quartic to be derived (the reduced form has no $x^3$ term).  This is easily done by putting $x = y + X_{Nf}$, where $X_{Nf}$ is equal to $( -b / 4a )$, and substituting for $x$ in the original quartic expression.  Euler's quartic solution showed that each of the roots of the reduced quartic is the sum of three square roots, $\pm\sqrt{r_1}\pm\sqrt{r_2}\pm\sqrt{r_3}$, where $r_1$ etc. can be derived directly from the three roots ($c_i$) of an underlying ``resolvent'' cubic.  A convenient reduced form (no quadratic term) of the resolvent cubic is:
\begin{eqnarray}	
	C(z) = z^3 - 3Iz + J = 0	\label{eq:resolventcubic}
\end{eqnarray}	
(with roots $c_1$, $c_2$, $c_3$), where the quartic invariants, $I$ and $J$, are given by
\begin{eqnarray}	
	I &=& 12ae - 3bd + c^2	\label{eq:Ieq} \\
	J &=& 72ace + 9bcd - 27ad^2 - 27eb^2 - 2c^3	\label{eq:Jeq}
\end{eqnarray}	

Furthermore, the invariants $I$ and $J$ can be combined to form the discriminant of the quartic equation $(4I^3 - J^2)$ and thus determine whether or not the resolvent cubic has three real roots (when $(4I^3 - J^2) > 0$) or one real root and one pair of complex conjugate roots (when $(4I^3 - J^2) < 0$).  In the quartic calculator, we use two different methods for finding $r_1$ etc. depending on whether not the resolvent cubic has only real or some complex roots.

\subsection{Real Roots}

If the resolvent cubic has three real roots (which we know must lead to a solution for our quartic equation of four real roots as one of those roots must lead to the real, physical value for $\mttwo$), then we can use a trigonometric method for calculating the roots as follows (as per \cite{Nickalls:1993aa}): we define an angle, $\theta$, such that

\begin{eqnarray}	
	\cos3\theta = -\frac{y_N}{h}  = -\frac{J}{\sqrt{4I^3}}  \label{eq:trigeq}
\end{eqnarray}	

Then the three roots of the reduced resolvent cubic are given by:
\begin{eqnarray}	
	 c_1 &=& 2\delta\cosTheta	\label{eq:trigalpha} \\
	c_2 &=& 2\delta\cos\left(\frac{2\pi}{3}+\theta\right) \label{eq:trigbeta} \\
	c_3 &=&  2\delta\cos\left(\frac{4\pi}{3}+\theta\right) \label{eq:triggamma}
\end{eqnarray}	

where $\delta = \sqrt{I}$.  The Euler values $r_1$, $r_2$, $r_3$, are then equal to:

\begin{eqnarray}	
	r_1 &=& \epsilon^2 +\frac{c_1}{12a} 	\label{eq:trigr1}\\
	r_2 &=& \epsilon^2 +\frac{c_2}{12a} 	\label{eq:trigr2}\\
	r_3 &=& \epsilon^2 +\frac{c_3}{12a} 	\label{eq:trigr3}
\end{eqnarray}	
where $\epsilon^2 = \frac{\left(3 b^2 - 8ac\right)}{48 a^2}$.  Note that as we know the quartic expression that leads to $\mttwo$ must have at least two real roots, we know that $r_1$, $r_2$ and $r_3$ must all be positive numbers (i.e. if any were negative, this would lead to four complex quartic roots).

\subsection{Real and Complex Roots}

If the resolvent cubic has one real root and one pair of complex conjugate roots then we must use a method which can handle complex numbers.  The real root of \eqref{eq:resolventcubic} is given by:
\begin{eqnarray}	
	c_1 = \sqrt[3]{\frac{1}{2}\left(-J + \sqrt{J^2 - 4I^3}\right)} + \sqrt[3]{\frac{1}{2}\left(-J - \sqrt{J^2 - 4I^3}\right)} 	\label{eq:complexalpha}
\end{eqnarray}	
and the pair of complex conjugate roots are given by:
\begin{eqnarray}	
	c_2 &=& -\frac{c_1}{2} + i\frac{\sqrt{3}}{2}\sqrt{{c_1}^2 - 4I}  \label{eq:complextbeta}\\
	c_3 &=& -\frac{c_1}{2} - i\frac{\sqrt{3}}{2}\sqrt{{c_1}^2 - 4I}	\label{eq:complexgamma}
\end{eqnarray}	

The three Euler values $r_1$, $r_2$, $r_3$, are then also given by \eqref{eq:trigr1}, \eqref{eq:trigr2} and \eqref{eq:trigr3} (but now we must remember that $r_2$ and $r_3$ are complex numbers).

\subsection{The Quartic Roots}

As has been mentioned previously, the four quartic roots are obtained from four of the eight possible permutations of $ \pm \sqrt{r_1} \pm \sqrt{r_2} \pm \sqrt{r_3}$.  Euler showed that for a given quartic equation, the sign of the product $ \sqrt{r_1} \sqrt{r_2} \sqrt{r_3}$ would be the same as the sign of $\left(\frac{-Y_{Nf'}}{a}\right)$ where $Y_{Nf'}$ is given by the following expression:
\begin{eqnarray}	
	Y_{Nf'} =\frac{b^3+8 a^2d-4abc}{8a^2}	\label{eq:YNFprime}
\end{eqnarray}	

As it is simplest to assume initially that all of the $\sqrt{r_i}$ are positive, we can then use the sign of $\left(\frac{-Y_{Nf'}}{a}\right)$, calculated from the quartic coefficients, to determine which four permutations are the ones that yield the correct roots of the original quartic equation \eqref{eq:generalquartic} (which we will call $X_1$, $X_2$, $X_3$, $X_4$). So we have the following two possible sets of results:
\mbox{}
If $\left(\frac{-Y_{Nf'}}{a}\right) < 0$
\begin{eqnarray}	
	X_1 = X_{Nf}  + \sqrt{r_1} + \sqrt{r_2} - \sqrt{r_3}	\label{eq:quarticrootoneneg}\\
	X_2 = X_{Nf}  + \sqrt{r_1} - \sqrt{r_2} + \sqrt{r_3}	\label{eq:quarticroottwoneg}\\
	X_3 = X_{Nf}  - \sqrt{r_1} + \sqrt{r_2} + \sqrt{r_3}	\label{eq:quarticrootthreeneg}\\
	X_4 = X_{Nf}  - \sqrt{r_1} - \sqrt{r_2} - \sqrt{r_3}	\label{eq:quarticrootfourneg}
\end{eqnarray}	
And when $\left(\frac{-Y_{Nf'}}{a}\right) > 0$
\begin{eqnarray}	
	X_1 = X_{Nf} - \sqrt{r_1} + \sqrt{r_2} - \sqrt{r_3}	\label{eq:quarticrootonepos}\\
	X_2 = X_{Nf} - \sqrt{r_1} - \sqrt{r_2} + \sqrt{r_3}	\label{eq:quarticroottwopos}\\
	X_3 = X_{Nf} + \sqrt{r_1} + \sqrt{r_2} + \sqrt{r_3}	\label{eq:quarticrootthreepos}\\
	X_4 = X_{Nf} + \sqrt{r_1} - \sqrt{r_2} - \sqrt{r_3}	\label{eq:quarticrootfourpos}
\end{eqnarray}	

Finally we note that in the case where we know that the above must give two real and one complex conjugate pair (i.e. when $(4I^3 - J^2) < 0$), then we know that the two real roots we want must be derived either from \eqref{eq:quarticrootthreeneg} and \eqref{eq:quarticrootfourneg} [or from \eqref{eq:quarticrootthreepos} and \eqref{eq:quarticrootfourpos} as appropriate] because only in these expressions will the imaginary parts of $\sqrt{r_2}$ and $\sqrt{r_3}$ cancel. Thus at this stage not only can we ignore \eqref{eq:quarticrootoneneg} and \eqref{eq:quarticroottwoneg} [or \eqref{eq:quarticrootonepos} and \eqref{eq:quarticroottwopos}], but we can also ignore the imaginary parts of \eqref{eq:quarticrootthreeneg} and \eqref{eq:quarticrootfourneg} [or \eqref{eq:quarticrootthreepos} and \eqref{eq:quarticrootfourpos}] because these must cancel. This is not just expedient, it also minimises any stability issues in the algorithm that might arise from errors in accurately calculating these complex numbers (as discussed previously).

Of course if we know we instead have four real roots (i.e. when $(4I^3 - J^2) > 0$), then we must evaluate all four roots as any one of them might be the one that leads to the minimum value for $\mttwo$.


\bibliography{paper}

\begin{thebibliography}{44}
\expandafter\ifx\csname natexlab\endcsname\relax\def\natexlab#1{#1}\fi
\expandafter\ifx\csname bibnamefont\endcsname\relax
  \def\bibnamefont#1{#1}\fi
\expandafter\ifx\csname bibfnamefont\endcsname\relax
  \def\bibfnamefont#1{#1}\fi
\expandafter\ifx\csname citenamefont\endcsname\relax
  \def\citenamefont#1{#1}\fi
\expandafter\ifx\csname url\endcsname\relax
  \def\url#1{\texttt{#1}}\fi
\expandafter\ifx\csname urlprefix\endcsname\relax\def\urlprefix{URL }\fi
\providecommand{\bibinfo}[2]{#2}
\providecommand{\eprint}[2][]{\url{#2}}

\bibitem[{\citenamefont{Lester and Summers}(1999)}]{Lester:1999tx}
\bibinfo{author}{\bibfnamefont{C.~G.} \bibnamefont{Lester}} \bibnamefont{and}
  \bibinfo{author}{\bibfnamefont{D.~J.} \bibnamefont{Summers}},
  \bibinfo{journal}{Phys. Lett.} \textbf{\bibinfo{volume}{B463}},
  \bibinfo{pages}{99} (\bibinfo{year}{1999}), \eprint{hep-ph/9906349}.

\bibitem[{\citenamefont{Allanach et~al.}(2000)\citenamefont{Allanach, Lester,
  Parker, and Webber}}]{Allanach:2000kt}
\bibinfo{author}{\bibfnamefont{B.~C.} \bibnamefont{Allanach}},
  \bibinfo{author}{\bibfnamefont{C.~G.} \bibnamefont{Lester}},
  \bibinfo{author}{\bibfnamefont{M.~A.} \bibnamefont{Parker}},
  \bibnamefont{and} \bibinfo{author}{\bibfnamefont{B.~R.}
  \bibnamefont{Webber}}, \bibinfo{journal}{JHEP} \textbf{\bibinfo{volume}{09}},
  \bibinfo{pages}{004} (\bibinfo{year}{2000}), \eprint{hep-ph/0007009}.

\bibitem[{\citenamefont{Barr et~al.}(2003{\natexlab{a}})\citenamefont{Barr,
  Lester, Parker, Allanach, and Richardson}}]{Barr:2002ex}
\bibinfo{author}{\bibfnamefont{A.~J.} \bibnamefont{Barr}},
  \bibinfo{author}{\bibfnamefont{C.~G.} \bibnamefont{Lester}},
  \bibinfo{author}{\bibfnamefont{M.~A.} \bibnamefont{Parker}},
  \bibinfo{author}{\bibfnamefont{B.~C.} \bibnamefont{Allanach}},
  \bibnamefont{and}
  \bibinfo{author}{\bibfnamefont{P.}~\bibnamefont{Richardson}},
  \bibinfo{journal}{JHEP} \textbf{\bibinfo{volume}{03}}, \bibinfo{pages}{045}
  (\bibinfo{year}{2003}{\natexlab{a}}), \eprint{hep-ph/0208214}.

\bibitem[{\citenamefont{Barr et~al.}(2003{\natexlab{b}})\citenamefont{Barr,
  Lester, and Stephens}}]{Barr:2003rg}
\bibinfo{author}{\bibfnamefont{A.}~\bibnamefont{Barr}},
  \bibinfo{author}{\bibfnamefont{C.}~\bibnamefont{Lester}}, \bibnamefont{and}
  \bibinfo{author}{\bibfnamefont{P.}~\bibnamefont{Stephens}},
  \bibinfo{journal}{J. Phys.} \textbf{\bibinfo{volume}{G29}},
  \bibinfo{pages}{2343} (\bibinfo{year}{2003}{\natexlab{b}}),
  \eprint{hep-ph/0304226}.

\bibitem[{\citenamefont{Lester and Barr}(2007)}]{Lester:2007fq}
\bibinfo{author}{\bibfnamefont{C.}~\bibnamefont{Lester}} \bibnamefont{and}
  \bibinfo{author}{\bibfnamefont{A.}~\bibnamefont{Barr}},
  \bibinfo{journal}{JHEP} \textbf{\bibinfo{volume}{12}}, \bibinfo{pages}{102}
  (\bibinfo{year}{2007}), \eprint{0708.1028}.

\bibitem[{\citenamefont{Cho et~al.}(2008{\natexlab{a}})\citenamefont{Cho, Choi,
  Kim, and Park}}]{Cho:2007qv}
\bibinfo{author}{\bibfnamefont{W.~S.} \bibnamefont{Cho}},
  \bibinfo{author}{\bibfnamefont{K.}~\bibnamefont{Choi}},
  \bibinfo{author}{\bibfnamefont{Y.~G.} \bibnamefont{Kim}}, \bibnamefont{and}
  \bibinfo{author}{\bibfnamefont{C.~B.} \bibnamefont{Park}},
  \bibinfo{journal}{Phys. Rev. Lett.} \textbf{\bibinfo{volume}{100}},
  \bibinfo{pages}{171801} (\bibinfo{year}{2008}{\natexlab{a}}),
  \eprint{0709.0288}.

\bibitem[{\citenamefont{Gripaios}(2008)}]{Gripaios:2007is}
\bibinfo{author}{\bibfnamefont{B.}~\bibnamefont{Gripaios}},
  \bibinfo{journal}{JHEP} \textbf{\bibinfo{volume}{02}}, \bibinfo{pages}{053}
  (\bibinfo{year}{2008}), \eprint{0709.2740}.

\bibitem[{\citenamefont{Barr et~al.}(2008{\natexlab{a}})\citenamefont{Barr,
  Gripaios, and Lester}}]{Barr:2007hy}
\bibinfo{author}{\bibfnamefont{A.~J.} \bibnamefont{Barr}},
  \bibinfo{author}{\bibfnamefont{B.}~\bibnamefont{Gripaios}}, \bibnamefont{and}
  \bibinfo{author}{\bibfnamefont{C.~G.} \bibnamefont{Lester}},
  \bibinfo{journal}{JHEP} \textbf{\bibinfo{volume}{02}}, \bibinfo{pages}{014}
  (\bibinfo{year}{2008}{\natexlab{a}}), \eprint{0711.4008}.

\bibitem[{\citenamefont{Cho et~al.}(2008{\natexlab{b}})\citenamefont{Cho, Choi,
  Kim, and Park}}]{Cho:2007dh}
\bibinfo{author}{\bibfnamefont{W.~S.} \bibnamefont{Cho}},
  \bibinfo{author}{\bibfnamefont{K.}~\bibnamefont{Choi}},
  \bibinfo{author}{\bibfnamefont{Y.~G.} \bibnamefont{Kim}}, \bibnamefont{and}
  \bibinfo{author}{\bibfnamefont{C.~B.} \bibnamefont{Park}},
  \bibinfo{journal}{JHEP} \textbf{\bibinfo{volume}{02}}, \bibinfo{pages}{035}
  (\bibinfo{year}{2008}{\natexlab{b}}), \eprint{0711.4526}.

\bibitem[{\citenamefont{Ross and Serna}(2008)}]{Ross:2007rm}
\bibinfo{author}{\bibfnamefont{G.~G.} \bibnamefont{Ross}} \bibnamefont{and}
  \bibinfo{author}{\bibfnamefont{M.}~\bibnamefont{Serna}},
  \bibinfo{journal}{Phys. Lett.} \textbf{\bibinfo{volume}{B665}},
  \bibinfo{pages}{212} (\bibinfo{year}{2008}), \eprint{0712.0943}.

\bibitem[{\citenamefont{Nojiri et~al.}(2008)\citenamefont{Nojiri, Polesello,
  and Tovey}}]{Nojiri:2007pq}
\bibinfo{author}{\bibfnamefont{M.~M.} \bibnamefont{Nojiri}},
  \bibinfo{author}{\bibfnamefont{G.}~\bibnamefont{Polesello}},
  \bibnamefont{and} \bibinfo{author}{\bibfnamefont{D.~R.} \bibnamefont{Tovey}},
  \bibinfo{journal}{JHEP} \textbf{\bibinfo{volume}{05}}, \bibinfo{pages}{014}
  (\bibinfo{year}{2008}), \eprint{0712.2718}.

\bibitem[{\citenamefont{Tovey}(2008)}]{Tovey:2008ui}
\bibinfo{author}{\bibfnamefont{D.~R.} \bibnamefont{Tovey}},
  \bibinfo{journal}{JHEP} \textbf{\bibinfo{volume}{04}}, \bibinfo{pages}{034}
  (\bibinfo{year}{2008}), \eprint{0802.2879}.

\bibitem[{\citenamefont{Cho et~al.}(2008{\natexlab{c}})\citenamefont{Cho, Choi,
  Kim, and Park}}]{Cho:2008cu}
\bibinfo{author}{\bibfnamefont{W.~S.} \bibnamefont{Cho}},
  \bibinfo{author}{\bibfnamefont{K.}~\bibnamefont{Choi}},
  \bibinfo{author}{\bibfnamefont{Y.~G.} \bibnamefont{Kim}}, \bibnamefont{and}
  \bibinfo{author}{\bibfnamefont{C.~B.} \bibnamefont{Park}},
  \bibinfo{journal}{Phys. Rev.} \textbf{\bibinfo{volume}{D78}},
  \bibinfo{pages}{034019} (\bibinfo{year}{2008}{\natexlab{c}}),
  \eprint{0804.2185}.

\bibitem[{\citenamefont{Serna}(2008)}]{Serna:2008zk}
\bibinfo{author}{\bibfnamefont{M.}~\bibnamefont{Serna}},
  \bibinfo{journal}{JHEP} \textbf{\bibinfo{volume}{06}}, \bibinfo{pages}{004}
  (\bibinfo{year}{2008}), \eprint{0804.3344}.

\bibitem[{\citenamefont{Barr et~al.}(2008{\natexlab{b}})\citenamefont{Barr,
  Ross, and Serna}}]{Barr:2008ba}
\bibinfo{author}{\bibfnamefont{A.~J.} \bibnamefont{Barr}},
  \bibinfo{author}{\bibfnamefont{G.~G.} \bibnamefont{Ross}}, \bibnamefont{and}
  \bibinfo{author}{\bibfnamefont{M.}~\bibnamefont{Serna}},
  \bibinfo{journal}{Phys. Rev.} \textbf{\bibinfo{volume}{D78}},
  \bibinfo{pages}{056006} (\bibinfo{year}{2008}{\natexlab{b}}),
  \eprint{0806.3224}.

\bibitem[{\citenamefont{Cho et~al.}(2009)\citenamefont{Cho, Choi, Kim, and
  Park}}]{Cho:2008tj}
\bibinfo{author}{\bibfnamefont{W.~S.} \bibnamefont{Cho}},
  \bibinfo{author}{\bibfnamefont{K.}~\bibnamefont{Choi}},
  \bibinfo{author}{\bibfnamefont{Y.~G.} \bibnamefont{Kim}}, \bibnamefont{and}
  \bibinfo{author}{\bibfnamefont{C.~B.} \bibnamefont{Park}},
  \bibinfo{journal}{Phys. Rev.} \textbf{\bibinfo{volume}{D79}},
  \bibinfo{pages}{031701} (\bibinfo{year}{2009}), \eprint{0810.4853}.

\bibitem[{\citenamefont{Burns et~al.}(2009)\citenamefont{Burns, Kong, Matchev,
  and Park}}]{Burns:2008va}
\bibinfo{author}{\bibfnamefont{M.}~\bibnamefont{Burns}},
  \bibinfo{author}{\bibfnamefont{K.}~\bibnamefont{Kong}},
  \bibinfo{author}{\bibfnamefont{K.~T.} \bibnamefont{Matchev}},
  \bibnamefont{and} \bibinfo{author}{\bibfnamefont{M.}~\bibnamefont{Park}},
  \bibinfo{journal}{JHEP} \textbf{\bibinfo{volume}{03}}, \bibinfo{pages}{143}
  (\bibinfo{year}{2009}), \eprint{0810.5576}.

\bibitem[{\citenamefont{Barr et~al.}(2009{\natexlab{a}})\citenamefont{Barr,
  Pinder, and Serna}}]{Barr:2008hv}
\bibinfo{author}{\bibfnamefont{A.~J.} \bibnamefont{Barr}},
  \bibinfo{author}{\bibfnamefont{A.}~\bibnamefont{Pinder}}, \bibnamefont{and}
  \bibinfo{author}{\bibfnamefont{M.}~\bibnamefont{Serna}},
  \bibinfo{journal}{Phys. Rev.} \textbf{\bibinfo{volume}{D79}},
  \bibinfo{pages}{074005} (\bibinfo{year}{2009}{\natexlab{a}}),
  \eprint{0811.2138}.

\bibitem[{\citenamefont{Barr et~al.}(2009{\natexlab{b}})\citenamefont{Barr,
  Gripaios, and Lester}}]{Barr:2009mx}
\bibinfo{author}{\bibfnamefont{A.~J.} \bibnamefont{Barr}},
  \bibinfo{author}{\bibfnamefont{B.}~\bibnamefont{Gripaios}}, \bibnamefont{and}
  \bibinfo{author}{\bibfnamefont{C.~G.} \bibnamefont{Lester}},
  \bibinfo{journal}{JHEP} \textbf{\bibinfo{volume}{07}}, \bibinfo{pages}{072}
  (\bibinfo{year}{2009}{\natexlab{b}}), \eprint{0902.4864}.

\bibitem[{\citenamefont{Barr et~al.}(2009{\natexlab{c}})\citenamefont{Barr,
  Gripaios, and Lester}}]{Barr:2009jv}
\bibinfo{author}{\bibfnamefont{A.~J.} \bibnamefont{Barr}},
  \bibinfo{author}{\bibfnamefont{B.}~\bibnamefont{Gripaios}}, \bibnamefont{and}
  \bibinfo{author}{\bibfnamefont{C.~G.} \bibnamefont{Lester}},
  \bibinfo{journal}{JHEP} \textbf{\bibinfo{volume}{11}}, \bibinfo{pages}{096}
  (\bibinfo{year}{2009}{\natexlab{c}}), \eprint{0908.3779}.

\bibitem[{\citenamefont{Polesello and Tovey}(2010)}]{Polesello:2009rn}
\bibinfo{author}{\bibfnamefont{G.}~\bibnamefont{Polesello}} \bibnamefont{and}
  \bibinfo{author}{\bibfnamefont{D.~R.} \bibnamefont{Tovey}},
  \bibinfo{journal}{JHEP} \textbf{\bibinfo{volume}{03}}, \bibinfo{pages}{030}
  (\bibinfo{year}{2010}), \eprint{0910.0174}.

\bibitem[{\citenamefont{Kim}(2010)}]{Kim:2009si}
\bibinfo{author}{\bibfnamefont{I.-W.} \bibnamefont{Kim}},
  \bibinfo{journal}{Phys. Rev. Lett.} \textbf{\bibinfo{volume}{104}},
  \bibinfo{pages}{081601} (\bibinfo{year}{2010}), \eprint{0910.1149}.

\bibitem[{\citenamefont{Konar et~al.}(2009{\natexlab{a}})\citenamefont{Konar,
  Kong, Matchev, and Park}}]{Konar:2009wn}
\bibinfo{author}{\bibfnamefont{P.}~\bibnamefont{Konar}},
  \bibinfo{author}{\bibfnamefont{K.}~\bibnamefont{Kong}},
  \bibinfo{author}{\bibfnamefont{K.~T.} \bibnamefont{Matchev}},
  \bibnamefont{and} \bibinfo{author}{\bibfnamefont{M.}~\bibnamefont{Park}}
  (\bibinfo{year}{2009}{\natexlab{a}}), \eprint{0910.3679}.

\bibitem[{\citenamefont{Konar et~al.}(2009{\natexlab{b}})\citenamefont{Konar,
  Kong, Matchev, and Park}}]{Konar:2009qr}
\bibinfo{author}{\bibfnamefont{P.}~\bibnamefont{Konar}},
  \bibinfo{author}{\bibfnamefont{K.}~\bibnamefont{Kong}},
  \bibinfo{author}{\bibfnamefont{K.~T.} \bibnamefont{Matchev}},
  \bibnamefont{and} \bibinfo{author}{\bibfnamefont{M.}~\bibnamefont{Park}}
  (\bibinfo{year}{2009}{\natexlab{b}}), \eprint{0911.4126}.

\bibitem[{\citenamefont{Baringer et~al.}(2011)\citenamefont{Baringer, Kong,
  McCaskey, and Noonan}}]{Baringer:2011nh}
\bibinfo{author}{\bibfnamefont{P.}~\bibnamefont{Baringer}},
  \bibinfo{author}{\bibfnamefont{K.}~\bibnamefont{Kong}},
  \bibinfo{author}{\bibfnamefont{M.}~\bibnamefont{McCaskey}}, \bibnamefont{and}
  \bibinfo{author}{\bibfnamefont{D.}~\bibnamefont{Noonan}},
  \bibinfo{journal}{JHEP} \textbf{\bibinfo{volume}{1110}}, \bibinfo{pages}{101}
  (\bibinfo{year}{2011}), \eprint{1109.1563}.

\bibitem[{\citenamefont{Aad et~al.}(2011)}]{daCosta:2011qk}
\bibinfo{author}{\bibfnamefont{G.}~\bibnamefont{Aad}} \bibnamefont{et~al.}
  (\bibinfo{collaboration}{Atlas Collaboration}), \bibinfo{journal}{Phys.Lett.}
  \textbf{\bibinfo{volume}{B701}}, \bibinfo{pages}{186} (\bibinfo{year}{2011}),
  \eprint{1102.5290}.

\bibitem[{\citenamefont{Chatrchyan
  et~al.}(2012{\natexlab{a}})}]{Collaboration:2012ida}
\bibinfo{author}{\bibfnamefont{S.}~\bibnamefont{Chatrchyan}}
  \bibnamefont{et~al.} (\bibinfo{collaboration}{CMS Collaboration})
  (\bibinfo{year}{2012}{\natexlab{a}}), \eprint{1207.1798}.

\bibitem[{\citenamefont{Aad et~al.}(2012)}]{c:2012gg}
\bibinfo{author}{\bibfnamefont{G.}~\bibnamefont{Aad}} \bibnamefont{et~al.}
  (\bibinfo{collaboration}{ATLAS Collaboration}) (\bibinfo{year}{2012}),
  \eprint{1208.2884}.

\bibitem[{\citenamefont{Chatrchyan et~al.}(2012{\natexlab{b}})}]{cm:2012jx}
\bibinfo{author}{\bibfnamefont{S.}~\bibnamefont{Chatrchyan}}
  \bibnamefont{et~al.} (\bibinfo{collaboration}{CMS Collaboration})
  (\bibinfo{year}{2012}{\natexlab{b}}), \eprint{1207.1798}.

\bibitem[{\citenamefont{Weber and Collaboration}(2012)}]{Weber:2012fa}
\bibinfo{author}{\bibfnamefont{H.}~\bibnamefont{Weber}} \bibnamefont{and}
  \bibinfo{author}{\bibfnamefont{f.~t.~C.} \bibnamefont{Collaboration}}
  (\bibinfo{collaboration}{CMS Collaboration}), \bibinfo{journal}{EPJ Web
  Conf.} \textbf{\bibinfo{volume}{28}}, \bibinfo{pages}{12021}
  (\bibinfo{year}{2012}), \eprint{1201.4659}.

\bibitem[{\citenamefont{Abazov et~al.}(2012)}]{d0000:2012qq}
\bibinfo{author}{\bibfnamefont{V.~M.} \bibnamefont{Abazov}}
  \bibnamefont{et~al.} (\bibinfo{collaboration}{D0 Collaboration})
  (\bibinfo{year}{2012}), \eprint{1207.1041}.

\bibitem[{\citenamefont{Barr et~al.}(2011)\citenamefont{Barr, French, Frost,
  and Lester}}]{Barr:2011he}
\bibinfo{author}{\bibfnamefont{A.~J.} \bibnamefont{Barr}},
  \bibinfo{author}{\bibfnamefont{S.~T.} \bibnamefont{French}},
  \bibinfo{author}{\bibfnamefont{J.~A.} \bibnamefont{Frost}}, \bibnamefont{and}
  \bibinfo{author}{\bibfnamefont{C.~G.} \bibnamefont{Lester}},
  \bibinfo{journal}{JHEP} \textbf{\bibinfo{volume}{1110}}, \bibinfo{pages}{080}
  (\bibinfo{year}{2011}), \eprint{1106.2322}.

\bibitem[{\citenamefont{Aaltonen et~al.}(2010)}]{Aaltonen:2009rm}
\bibinfo{author}{\bibfnamefont{T.}~\bibnamefont{Aaltonen}} \bibnamefont{et~al.}
  (\bibinfo{collaboration}{CDF}), \bibinfo{journal}{Phys. Rev.}
  \textbf{\bibinfo{volume}{D81}}, \bibinfo{pages}{031102}
  (\bibinfo{year}{2010}), \eprint{0911.2956}.

\bibitem[{\citenamefont{Cheng and Han}(2008)}]{Cheng:2008hk}
\bibinfo{author}{\bibfnamefont{H.-C.} \bibnamefont{Cheng}} \bibnamefont{and}
  \bibinfo{author}{\bibfnamefont{Z.}~\bibnamefont{Han}},
  \bibinfo{journal}{JHEP} \textbf{\bibinfo{volume}{12}}, \bibinfo{pages}{063}
  (\bibinfo{year}{2008}), \eprint{0810.5178}.

\bibitem[{\citenamefont{Barr and Lester}()}]{oxbridgeStransverseMassLibrary}
\bibinfo{author}{\bibfnamefont{A.~J.} \bibnamefont{Barr}} \bibnamefont{and}
  \bibinfo{author}{\bibfnamefont{C.~G.} \bibnamefont{Lester}},
  \emph{\bibinfo{title}{Oxbridge stransverse mass library}},
  \bibinfo{note}{\url{http://www.hep.phy.cam.ac.uk/~lester/mt2/index.html}}.

\bibitem[{\citenamefont{Cheng et~al.}()\citenamefont{Cheng, Engelhardt, Gunion,
  Han, Marandella, and McElrath}}]{ucdWimpmassLibrary}
\bibinfo{author}{\bibfnamefont{H.-C.} \bibnamefont{Cheng}},
  \bibinfo{author}{\bibfnamefont{D.}~\bibnamefont{Engelhardt}},
  \bibinfo{author}{\bibfnamefont{J.~F.} \bibnamefont{Gunion}},
  \bibinfo{author}{\bibfnamefont{Z.}~\bibnamefont{Han}},
  \bibinfo{author}{\bibfnamefont{G.}~\bibnamefont{Marandella}},
  \bibnamefont{and} \bibinfo{author}{\bibfnamefont{B.}~\bibnamefont{McElrath}},
  \emph{\bibinfo{title}{{\tt WIMPMASS} library}},
  \bibinfo{note}{\url{http://particle.physics.ucdavis.edu/hefti/projects/doku.%
php?id=wimpmass}}.

\bibitem[{\citenamefont{Lester}(2011)}]{Lester:2011nj}
\bibinfo{author}{\bibfnamefont{C.~G.} \bibnamefont{Lester}},
  \bibinfo{journal}{JHEP} \textbf{\bibinfo{volume}{1105}}, \bibinfo{pages}{076}
  (\bibinfo{year}{2011}), \eprint{1103.5682}.

\bibitem[{\citenamefont{Nickalls}(2009)}]{Nickalls:2009aa}
\bibinfo{author}{\bibfnamefont{R.~W.~D.} \bibnamefont{Nickalls}},
  \bibinfo{journal}{The Mathematical Gazette} \textbf{\bibinfo{volume}{93}},
  \bibinfo{pages}{66} (\bibinfo{year}{2009}),
  \urlprefix\url{http://www.nickalls.org/dick/papers/maths/quartic2009.pdf}.

\bibitem[{\citenamefont{Cardano}(1993)}]{CardanoArsMagna}
\bibinfo{author}{\bibfnamefont{G.}~\bibnamefont{Cardano}},
  \emph{\bibinfo{title}{Ars magna or The Rules of Algebra}},
  \bibinfo{number}{0486678113} (\bibinfo{publisher}{Dover},
  \bibinfo{year}{1993}).

\bibitem[{\citenamefont{Shmakov}(2011)}]{Shmakov:2011aa}
\bibinfo{author}{\bibfnamefont{S.~L.} \bibnamefont{Shmakov}},
  \bibinfo{journal}{IJPAM} \textbf{\bibinfo{volume}{71-2}},
  \bibinfo{pages}{251} (\bibinfo{year}{2011}), \eprint{1314-3395}.

\bibitem[{\citenamefont{Herbison-Evans}(2011)}]{HerbisonEvans:2011aa}
\bibinfo{author}{\bibfnamefont{D.}~\bibnamefont{Herbison-Evans}},
  \bibinfo{type}{Tech. Rep.} \bibinfo{number}{TR94-487},
  \bibinfo{institution}{University of Sydney}, \bibinfo{address}{Sydney,
  Australia} (\bibinfo{year}{2011}),
  \urlprefix\url{http://donhe.topcities.com/pubs/solving.html}.

\bibitem[{\citenamefont{Bell}(2008)}]{Bell:2008le}
\bibinfo{author}{\bibfnamefont{J.}~\bibnamefont{Bell}} (\bibinfo{year}{2008}),
  \bibinfo{note}{an English translation of Euler's \textit {De formis radicum
  aequationum cuiusque ordinis coniectatio} (1733) (E30)},
  \eprint{0806.1927v1}.

\bibitem[{\citenamefont{Euler}(2006)}]{EulerElements}
\bibinfo{author}{\bibfnamefont{L.}~\bibnamefont{Euler}},
  \emph{\bibinfo{title}{Elements of Algebra}} (\bibinfo{publisher}{Tarquin
  Publications}, \bibinfo{year}{2006}), \bibinfo{note}{[English translation of
  Euler 1770 (E387)]}.

\bibitem[{\citenamefont{Nickalls}(1993)}]{Nickalls:1993aa}
\bibinfo{author}{\bibfnamefont{R.~W.~D.} \bibnamefont{Nickalls}},
  \bibinfo{journal}{The Mathematical Gazette} \textbf{\bibinfo{volume}{77}},
  \bibinfo{pages}{354} (\bibinfo{year}{1993}),
  \urlprefix\url{http://www.nickalls.org/dick/papers/maths/cubic1993.pdf}.

\end{thebibliography}
\end{document}